\documentclass[12pt]{article}
\usepackage[numbers]{natbib}
\usepackage{a4}
\usepackage{amsmath}
\usepackage{amssymb}
\usepackage{latexsym}
\usepackage{amssymb}
\usepackage{epsfig}
\usepackage{natbib}
\usepackage{bm}
\usepackage{color}
\def \d{{\mathrm{d}}}

\def \pd{\partial}

\def \Bs{\boldsymbol{s}}

\def \BR{{\boldsymbol{R}}}
\def \BV{{\boldsymbol{V}}}

\def \BJ{{\boldsymbol{J}}}

\def \rr{{\boldsymbol{r}}}

\def \BB{{\boldsymbol{B}}}
\def \BE{{\boldsymbol{E}}}
\def \BH{{\boldsymbol{H}}}
\def \BD{{\boldsymbol{D}}}
\def \BA{{\boldsymbol{A}}}

\def \tR{{t_{\text{R}}}}

\begin{document}
%%%%%%%%%%%%%%%%%%%%%%%%%%%%%%%%%%%%%%%%%%%%%%%%%%%%%%%%%%%%%%%%%%%%%
\title{{\bf 
Green functions and propagation in the Bopp-Podolsky electrodynamics
}}
\author{
Markus Lazar~$^\text{}$\footnote{
%%Corresponding author. 
%Tel.:+49(0)6151/163686; Fax.: +49(0)6151/163681.
{\it E-mail address:} lazar@fkp.tu-darmstadt.de (M.~Lazar).
%%\newline
%%Tel.:+49(0)6151/163686; Fax.: +49(0)6151/163681.
%%\newline
%%Tel.:~+49(0)6151/163686;~Fax.:~+49(0)6151/163681.
}
\\ \\
%${}^\text{a}$ 
%%        Heisenberg Research Group,\\
        Department of Physics,\\
        Darmstadt University of Technology,\\
        Hochschulstr. 6,\\      
        D-64289 Darmstadt, Germany\\
%%${}^\text{b}$ 
%%Department of Physics,\\
%%Michigan Technological University,\\
%%Houghton, MI 49931, USA
%%%${}^\text{b}$ 
}

\date{}    
\maketitle
%%%%%%%%%%%%%%%%%%%%%%%%%%%%%%%%%%%%%%%%%%%%%%%%%%%%%%%%%%%%%%%%%%%%%%%%%%%%%

{\sf Wave Motion~{\bf 91} (2019), 102388; 
https://doi.org/10.1016/j.wavemoti.2019.102388}
\\

%%%%\newpage
\begin{abstract}
In this paper,  we investigate the so-called Bopp-Podolsky electrodynamics.
The Bopp-Podolsky electrodynamics is a prototypical
gradient field theory  with weak nonlocality in space and time.  
The Bopp-Podolsky electrodynamics is a Lorentz and gauge invariant generalization of the 
Maxwell electrodynamics.
We derive the retarded Green functions, first derivatives of the retarded Green functions,
retarded potentials, retarded electromagnetic field strengths, 
generalized Li{\'e}nard-Wiechert potentials and the corresponding electromagnetic field strengths 
 in the framework of the Bopp-Podolsky electrodynamics for three, two and one spatial  dimensions.
We investigate the behaviour of these electromagnetic fields in the neighbourhood of the light cone.  
In the Bopp-Podolsky electrodynamics, the retarded Green functions and their first
derivatives show fast decreasing oscillations inside the forward light cone. 
\\ 

\noindent
{\bf Keywords:} Bopp-Podolsky electrodynamics; Green function; 
propagation; retardation; retarded potentials; Li{\'e}nard-Wiechert potentials\\
\end{abstract}

\section{Introduction}
Generalized continuum theories such as gradient theories and nonlocal theories
are exciting and challenging research fields 
in physics, applied mathematics, material science and engineering science (see, e.g., \citep{Bopp,Bopp2,Podolsky,PS,Mindlin64,KK66,Maugin79,Eringen02,LM05,AL09}). 
Gradient theories and nonlocal theories possess characteristic internal length scales 
in order to describe size-effects. 
Such generalized continuum theories are able to provide a regularization 
of the singularities present in classical continuum theories
which are not valid at short distances.
Generalized continuum theories are continuum theories
valid at small scales.

In physics, an important and useful gradient theory is the so-called Bopp-Podolsky electrodynamics,
which is the gradient theory of electrodynamics containing one length scale parameter,
$\ell$, the so-called Bopp-Podolsky parameter. 
\citet{Bopp} and \citet{Podolsky} 
have proposed such a gradient theory representing a classical  generalization of the 
Maxwell electrodynamics towards a generalized electrodynamics with linear field equations of fourth order 
in order to avoid singularities in the electromagnetic fields and to have a finite and positive self-energy of point charges
(see also~\citep{PS,Iwan,Lande3}). 
Due to its simplicity,
the Bopp-Podolsky theory can be considered as the prototype of a gradient theory.
Therefore, the Bopp-Podolsky electrodynamics represents the simplest, physical gradient field theory with weak nonlocality in space and time. 

Nowadays there is a renewed interest in the Bopp-Podolsky electrodynamics (e.g., \citep{Zayats,Perlick2015}), in particular to solve the 
long-outstanding problem of the electromagnetic self-force of a charged particle present in the classical Maxwell electrodynamics
which goes back to Lorentz, Abraham and Dirac trying to formulate a classical theory of the electron. 
The equation of motion in  the classical theory of the electron, often called
the Lorentz-Dirac equation,
is of third order in the time-derivative of the particle position, and as
result it shows unphysical behaviour such as run-away solutions and
pre-acceleration (see, e.g., the books by~\citet{Rohrlich} and \citet{Spohn}). 
Therefore, the classical Maxwell electrodynamics in vacuum does not lead to a
consistent equation of  motion of charged point particles and a generalized
electrodynamics could solve this problem.
%%In particular, a mathematically consistent theory of classical charged point particles
%%is of high relevance in accelerator physics to describe beams in terms of classical point particles.
%and not in terms of extended classical charged distributions or quantum particles.

In the static case,
gradient electrostatics with generalized Coulomb law was given by~\citet{Bopp,Podolsky,Lande3}
and gradient magnetostatics  including the generalized Biot-Savart law was given by~\citet{Lazar2014}.
Such generalized electrostatics and generalized magnetostatics have a physical meaning 
if the classical electric and magnetic fields are recovered in the limit $\ell\rightarrow 0$.
In gradient electrostatics, for a point charge the electric potential is finite and non-singular,
but the electric field strength is finite and discontinuous at the position of the point charge.

The Bopp-Podolsky theory has many interesting features.
It solves the problem of infinite self-energy in the electrostatic case, and 
it gives the correct expression for the self-force of charged particles at 
short distances eliminating the singularity when $r\rightarrow 0$
as shown by ~\citet{Frenkel,Zayats,Perlick2015}. 
In this manner, the Bopp-Podolsky electrodynamics is free of classical divergences. 
Using the Bopp-Podolsky electrodynamics, \citet{Frenkel} solved the 
so-called $4/3$ problem of the electromagnetic mass in the Abraham-Lorentz theory,
and \citet{Frenkel99} eliminated runaway solutions from the Lorentz-Dirac equation of motion.
These features allow experiments that could test the generalized
electrodynamics as a viable effective field theory~(e.g., \citep{Cuzi}) and 
the Bopp-Podolsky electrodynamics offers the possibilities of the physical modeling at small scales.
\citet{Iwan,Kvasnica} and \citet{Cuzi} argued that the Bopp-Podolsky length scale parameter $\ell$ 
is in the order of $\sim 10^{-15}$\, m, that means femtometre (fm), 
or even smaller.
From the mathematical point of view, the length parameter $\ell$ 
plays the role of the regularization parameter in the Bopp-Podolsky electrodynamics.
This length scale is associated to the massive mode, $m_{\text{BP}}$, 
of the Bopp-Podolsky electrodynamics through 
$m_{\text{BP}}=\hbar/(c\ell)$. 
Moreover, it is interesting to note 
that the Bopp-Podolsky electrodynamics is the
only linear generalization of the Maxwell electrodynamics 
whose Lagrangian, containing  
second order derivatives of the electromagnetic gauge potentials, 
is both Lorentz and $U(1)$-gauge invariant~\citep{Cuz}.

The Bopp-Podolsky electrodynamics is akin to the Pauli-Villars 
regularization procedure used in quantum electrodynamics 
(see, e.g.,~\citep{Kvasnica,Kvasnica2,Zayats,Ji}). 
Therefore, the Bopp-Podolsky electrodynamics provides a regularization 
of the Maxwell electrodynamics based on higher order partial differential equations.
On the other hand, \citet{Santos2011} analyzed the wave propagation in the vacuum of 
the Bopp-Podolsky electrodynamics and two kinds of waves were found: 
the classical non-dispersive wave of the Maxwell electrodynamics, and a dispersive wave reminiscent of wave propagation 
in a collisionless plasma  with plasma (angular) frequency $\omega_p=c/\ell$, described by a Klein-Gordon equation. 

In the Maxwell electrodynamics, quantities like the 
retarded potentials, the retarded electromagnetic field strengths, the
Li{\'e}nard-Wiechert potentials and the electromagnetic field strengths
in the Li{\'e}nard-Wiechert form are the basic fields and quantities for the classical electromagnetic radiation 
(see, e.g., \citep{Jackson,HM,Smith}). 
In particular, the  Li{\'e}nard-Wiechert form of the electromagnetic field strengths 
is important for the calculation of the self-force of a charged point particle.
In the Bopp-Podolsky electrodynamics, only a little is known for such fields 
necessary for the electromagnetic radiation and radiation reaction in the generalized electrodynamics of
Bopp and Podolsky and their behaviour on the light cone (see, e.g., \citep{Lande,Zayats,Perlick2015}).
Only, 
the three-dimensional generalized Li{\'e}nard-Wiechert potentials were given 
by~\citet{Lande} and the corresponding three-dimensional
electromagnetic fields of a point charge have been recently given 
by~\citet{Perlick2015}, for the first time.
The aim of the present work is to close this gap and to 
give a systematic derivation and presentation of 
all important quantities in 
three, two and one spatial  dimensions (3D, 2D, 1D).
In particular, 
this work gives, for the first time, the 
analytical expressions for the retarded potentials and 
retarded electromagnetic fields in 2D and 1D,
and for the  generalized Li{\'e}nard-Wiechert potentials and corresponding electromagnetic fields 
of a non-uniformly moving charge in 2D and 1D in the framework of the Bopp-Podolsky electrodynamics.
This completes the library of
all important field solutions
needed in the Bopp-Podolsky electrodynamics in 3D, 2D, and 1D, 
which is a necessary step towards completing the study of the Bopp-Podolsky electrodynamics. 
In particular, we investigate the behaviour of these fields near and on the light cone.

The purpose of this paper is to add relevant results of the Green functions, retardation and wave propagation in the Bopp-Podolsky electrodynamics. 
In Section~\ref{sec2}, we review the basic equations 
of the Bopp-Podolsky electrodynamics.
In Section~\ref{sec3}, we give 
a systematic derivation and 
collection of the (dynamical) Bopp-Podolsky Green
function and its first derivatives in 3D, 2D and 1D in the framework of
generalized functions.
The retarded potentials and retarded electromagnetic field strengths are given
in Section~\ref{sec4} for 3D, 2D and 1D. 
In Section~\ref{sec5}, we present the generalized Li{\'e}nard-Wiechert potentials and electromagnetic field strengths
in generalized Li{\'e}nard-Wiechert form. 
The paper closes with the conclusion in Section~\ref{sec6}.

\section{Basic framework of the Bopp-Podolsky electrodynamics}
\label{sec2}

In the Bopp-Podolsky electrodynamics~\citep{Bopp,Podolsky}, the electromagnetic fields are described by the Lagrangian density
\begin{align}
\label{L-BP}
{\cal L_{\text {BP}}}&=
\frac{\varepsilon_0}{2}\, 
\Big(\bm E \cdot \bm E 
+\ell^2 \nabla \bm E :\nabla \bm E 
-\frac{\ell^2}{c^2}\, \pd_t \bm E \cdot \pd_t \bm E \Big)\nonumber\\
&\ 
-\frac{1}{2\mu_0 }\, 
\Big(\bm B \cdot \bm B 
+\ell^2 \nabla \bm B :\nabla \bm B 
-\frac{\ell^2}{c^2}\, \pd_t \bm B \cdot \pd_t \bm B \Big)
-\rho\phi+\bm J\cdot \bm A\,,
\end{align}
with the notation $ \nabla \bm E :\nabla \bm E =\pd_j E_i \pd_j E_i$ and
$ \bm E \cdot\bm E =E_i E_i$.  
Eq.~\eqref{L-BP} corresponds to Bopp's form of the Lagrangian~\citep{Bopp}. 
Here 
$\phi$ and $\BA$ are the electromagnetic gauge potentials,
$\BE$ is the electric field strength vector,
$\BB$ is the magnetic field strength vector,
$\rho$ is the electric charge density, and 
$\BJ$ is the electric current density vector. 
$\varepsilon_0$ is the electric constant 
and 
$\mu_0$ is the magnetic constant 
(also called permittivity of vacuum
and permeability of vacuum, respectively).
The speed of light in vacuum is given by
\begin{align}
c=\frac{1}{\sqrt{\varepsilon_0\mu_0}}\,.
\end{align}
Moreover, $\ell$ is the characteristic length scale parameter in the Bopp-Podolsky electrodynamics,
$\pd_t$ denotes the differentiation with respect to the time $t$ and
$\nabla$ is the Nabla operator.
From the mathematical point of view, the characteristic length parameter~$\ell$
plays the role of a regularization parameter in the Bopp-Podolsky theory.
In addition to the classical terms, first spatial- and time-derivatives of the electromagnetic field strengths   ($\BE$, $\BB$) 
multiplied by the characteristic length $\ell$ and a characteristic time $T=\ell/c$, respectively, appear
in Eq.~\eqref{L-BP} which describe a weak nonlocality in space and time.
The limit $\ell\rightarrow 0$ is the limit from the Bopp-Podolsky electrodynamics to 
the Maxwell electrodynamics.

The electromagnetic field strengths ($\BE$, $\BB$)
can be expressed in terms
of the electromagnetic gauge potentials (scalar potential $\phi$, vector potential $\BA$)
\begin{align} 
\label{E}
\BE&=-\nabla \phi-\pd_t \bm A\,,\\
\label{B}
\BB&=\nabla\times \bm A\,.
\end{align}
Due to their mathematical structure, the electromagnetic field strengths~\eqref{E} and \eqref{B}
satisfy the two electromagnetic Bianchi identities (or electromagnetic  compatibility conditions)
\begin{align}
\label{BI-1}
\nabla\times\BE+\pd_t\BB&=0\,,\\
\label{BI-2}
\nabla\cdot \BB&=0\,,
\end{align}
which are known as homogeneous Maxwell equations.

The Euler-Lagrange equations of the Lagrangian~\eqref{L-BP} with respect to the scalar potential $\phi$ and the 
vector potential $\bm A$ give the electromagnetic field equations
\begin{align}
\label{EL-1}
&\big[1+\ell^2 \square \big]\,
\nabla\cdot \bm E=\frac{1}{\varepsilon_0}\,\rho\,,\\
\label{EL-2}
&\big[1+\ell^2 \square \big]
\Big(
\nabla\times\bm B-\frac{1}{c^2}\,\pd_t\bm E\Big)=\mu_0\,\BJ\,,
\end{align}
respectively. The d'Alembert operator is defined as 
\begin{align}
\square:=\frac{1}{c^2}\,\pd_{tt}-\Delta\,,
%%\qquad\text{with}\quad\Delta=\nabla^2\, .
\end{align}
where $\Delta$ is the Laplace operator.
Eqs.~\eqref{EL-1} and \eqref{EL-2} represent the generalized inhomogeneous
Maxwell equations in the Bopp-Podolsky electrodynamics.  
In addition, the electric current density vector 
and the electric charge density
fulfill the continuity equation
\begin{align}
\label{CE}
&\nabla\cdot \BJ+\pd_t\rho=0\,.
\end{align}

If we use the variational derivative with respect to 
the electromagnetic fields ($\bm E$, $\bm B$),
we obtain the constitutive relations in the Bopp-Podolsky electrodynamics for the response quantities
($\bm D$, $\bm H$) in vacuum 
\begin{align}
\label{CE1}
\BD&:=\frac{\delta{\cal L_{\text {BP}}}}{\delta \BE}=\varepsilon_0\, \big[1+\ell^2 \square \big]\BE\,,\\
\label{CE2}
\BH&:=-\frac{\delta{\cal L_{\text {BP}}}}{\delta \BB}=\frac{1}{\mu_0}\,\big[1+\ell^2 \square \big] \BB\,,
\end{align}
where $\BD$ is the electric displacement vector (electric excitation), 
$\BH$ is the magnetic excitation vector.
The second terms in Eqs.~\eqref{CE1} and \eqref{CE2} describe the polarization
of the vacuum present in the Bopp-Podolsky electrodynamics. 
The vacuum in the Bopp-Podolsky electrodynamics is a 
classical vacuum plus vacuum polarization that behaves like a 
plasma-like vacuum~\citep{Santos2011}. 

Using the constitutive relations~\eqref{CE1} and \eqref{CE2}, 
the Euler-Lagrange equations~\eqref{EL-1} and \eqref{EL-2} 
can be rewritten in the form of inhomogeneous Maxwell equations
\begin{align}
\label{ME-inh}
\nabla\cdot \BD&=\rho\,,\\
\label{ME-inh2}
 \nabla\times\BH-\pd_t\BD&=\BJ\,. 
\end{align}

From Eqs.~(\ref{EL-1}) and (\ref{EL-2}),
inhomogeneous Bopp-Podolsky  equations, being partial differential
equations of fourth order, follow for the electromagnetic field strengths
\begin{align}
\label{E-w}
\big[1+\ell^2 \square \big]\square\,\BE&=-\frac{1}{\varepsilon_0}\Big(\nabla\rho+\frac{1}{c^2}\, \pd_t\BJ\Big)\,,\\
%%\end{align}
%%and
%%\begin{align}
\label{B-w}
\big[1+\ell^2 \square \big]\square\,\BB&=\mu_0\,\nabla\times\BJ\,.
\end{align}
Using the generalized Lorentz gauge condition~\citep{GP}
\begin{align}
\label{LG}
\big[1+\ell^2 \square \big]
\left(\frac{1}{c^2}\,\pd_t \phi+ \nabla \cdot \bm A\right)=0\,,
\end{align}
the electromagnetic gauge potentials fulfill the following 
inhomogeneous Bopp-Podolsky equations 
\begin{align}
\label{phi-w}
\big[1+\ell^2 \square \big]\square\,\phi&=\frac{1}{\varepsilon_0}\, \rho\,,\\
%%\end{align}
%%and
%%\begin{align}
\label{A-w}
\big[1+\ell^2 \square \big]\square\,\bm A&=\mu_0\, \BJ\,.
\end{align}
Note  that the generalized Lorentz gauge condition~\eqref{LG}
is as natural in the Bopp-Podolsky electrodynamics as the
Lorentz gauge condition is in the Maxwell electrodynamics~\citep{GP}.
As shown by~\citet{GP},  the usual Lorentz gauge condition,
$\frac{1}{c^2}\,\pd_t \phi+ \nabla \cdot \bm A=0$, does not satisfy the necessary requirements for a consistent gauge
in the Bopp-Podolsky electrodynamics: 
it does not fix the gauge, it is not preserved by the
equations of motion, and it is not attainable.   
The generalized Lorentz gauge condition is also necessary in the quantization
of the Bopp-Podolsky electrodynamics leading to a generalized quantum electrodynamics~\citep{Bufalo11,Bufalo12}.
\citet{Bufalo11} found that in such a generalized quantum electrodynamics, using the one-loop approximation,
the electron self-energy and the vertex function are both ultraviolet finite.

\section{Green function of the Bopp-Podolsky equation}
\label{sec3}

The Bopp-Podolsky electrodynamics is a linear theory with partial differential
equations of fourth order.
Therefore,
the powerful method of Green functions (fundamental solutions) can be used to construct exact analytical solutions.

The Green function $G^{\rm BP}$ of the Bopp-Podolsky equation, which is a partial differential equation of fourth order,
is defined by 
\begin{align}
\label{BPE}
\big[1+\ell^2 \square \big]\square\, G^{\rm BP}(\bm R, \tau)=\delta(\tau)\delta(\bm R)\,,
%%\quad \text{for}\quad \tau>0
\end{align}
where $\tau=t-t'$, $\bm R= \bm r-\bm r'$ and $\delta$ is the Dirac $\delta$-function. 
Therefore, the Green function,  $G^{\rm BP}$, is the fundamental solution of the linear hyperbolic differential operator of fourth order,
$[1+\ell^2 \square \big]\square$, in the sense of Schwartz' distributions (or generalized functions)~\citep{Schwartz}.
Because we are only interested in the retarded Green function, the causality constraint must be fulfilled
\begin{align}
\label{CC}
G^{\rm BP}(\bm R, \tau)=0\qquad \text{for}\quad \tau<0\,.
\end{align}
As always for hyperbolic operators, the Green function $G^{\rm BP}(\bm R, \tau)$ is the only fundamental solution of the (hyperbolic) Bopp-Podolsky operator
with support in the half-space $\tau\ge 0$ (see, e.g., \citep{Hoermander}).

The Bopp-Podolsky equation~(\ref{BPE}) can be written as an equivalent  system of partial differential equations of second order 
\begin{align}
\label{BPE-2}
\big [1+\ell^2 \square \big] G^{\rm BP}(\bm R, \tau)&=G^\square(\bm R, \tau)\, ,\\
\label{BPE-3}
\square\, G^{\rm BP}(\bm R, \tau)&=G^{\rm KG}(\bm R, \tau)\, ,\\
\label{wave}
\square\, G^\square(\bm R, \tau)&=\delta(\tau)\delta(\bm R) \,,\\
\label{KGE}
\big [1+\ell^2 \square \big] G^{\rm KG}(\bm R, \tau)&=\delta(\tau)\delta(\bm R)\,,
\end{align}
where $G^\square$ is the Green function of the d'Alembert equation~\eqref{wave} and 
$G^{\rm KG}$ is the Green function of the Klein-Gordon equation~\eqref{KGE}.
It can be seen that the Bopp-Podolsky equation~(\ref{BPE}) is a Klein-Gordon-d'Alembert equation.
Finally, the Green function $G^{\rm BP}$ of the Bopp-Podolsky equation can be written in terms of the Green function $G^\square$ 
of the d'Alembert equation and the Green function $G^{\rm KG}$ of the Klein-Gordon equation (see also \citep{Lazar2010})
\begin{align}
\label{BPG}
G^{\rm BP}=G^\square-\ell^2 G^{\rm KG}\,,
\end{align}
or in the (formal) operator notation using the partial fraction decomposition 
\begin{align}
\label{Deco-op}
\big[\big(1+\ell^2 \square\big)\square\big]^{-1}=\square^{-1}-\ell^2\big[1+\ell^2 \square\big]^{-1}\,.
\end{align}
Using Eq.~\eqref{BPG}, the Green function of the Bopp-Podolsky equation can be
derived by means of the expressions of 
the Green function of the d'Alembert equation (see, e.g., \citep{Barton,Kanwal,Wl,Zauderer}) and
the Green function of the Klein-Gordon equation  (see, e.g., \citep{Iwan,Zauderer,Poly}).
Therefore, the Bopp-Podolsky field is a superposition of the Maxwell field
and the Klein-Gordon field. 
On the other hand, the Green function of the Bopp-Podolsky equation can be written as
convolution of the Green function of the d'Alembert operator and 
the Green function of the Klein-Gordon operator
\begin{align}
\label{GBP-conv}
G^{\rm BP}=G^\square*G^{\rm KG}\,,
\end{align}
satisfying Eqs.~\eqref{BPE}, \eqref{BPE-2} and \eqref{BPE-3}.
The symbol $*$ denotes the convolution in space and time. 
It can be seen in Eq.~\eqref{GBP-conv} that the Green function $G^{\rm KG}$ of the Klein-Gordon operator plays the role of the 
regularization function in the Bopp-Podolsky electrodynamics, regularizing the  Green function $G^\square$ 
of the d'Alembert operator towards the Green function $G^{\rm BP}$ of the Bopp-Podolsky operator.
On the other hand, the limit of $G^{\rm BP}$ as $\ell$ tends to zero reads (see Eq.~\eqref{BPG})
\begin{align}
\label{lim-BPG}
\lim_{\ell \to 0}
G^{\rm BP}=G^\square\,.
\end{align}

In this work,  we only consider the retarded Green functions which are zero for $\tau<0$.

\subsection{3D Green functions}

The three-dimensional Green functions (fundamental solutions) of the 
wave (d'Alembert) operator~\eqref{wave}, the Klein-Gordon operator~\eqref{KGE}
and the Bopp-Podolsky (Klein-Gordon-d'Alembert) operator are the
(generalized) functions ($\tau>0$):
\begin{align}
\label{G-w-3d}
G_{(3)}^\square(\bm R, \tau)&=\frac{1}{4\pi R}\,\delta\big(\tau-R/c\big)\,,\\
\label{G-KG-3d}
G_{(3)}^{\rm KG}(\bm R, \tau)
&=\frac{1}{4\pi\ell^2}\,\bigg[
\frac{1}{R}\,\delta\big(\tau-R/c\big)
-\frac{c}{\ell}\,
\frac{H\big(c\tau-R\big)}{\sqrt{c^2 \tau^2-R^2}}\, 
J_1 \bigg( \frac{\sqrt{c^2 \tau^2-R^2}}{\ell}\bigg)\bigg]\,,\\
%%\end{align}
%%where $H$ is the Heaviside step function.
\label{G-BP-3d}
G_{(3)}^{\rm BP}(\bm R, \tau)
&=\frac{c}{4\pi\ell}\,
\frac{H\big(c\tau-R\big)}{\sqrt{c^2 \tau^2-R^2}}\, 
J_1 \bigg(\frac{\sqrt{c^2 \tau^2-R^2}}{\ell}\bigg)\,,
\end{align}
where $R=\sqrt{(x-x')^2+(y-y')^2+(z-z')^2}$, $H$ is the Heaviside step function 
and $J_1$ is the Bessel function of the first kind of order one. 
Eq.~\eqref{G-BP-3d} is obtained from Eq.~\eqref{BPG} using the Green functions~\eqref{G-w-3d} and \eqref{G-KG-3d}.
The Green function~\eqref{G-BP-3d} is in agreement with the expression given earlier in \citep{Hoehler,Frenkel,Frenkel99}.

\begin{figure}[t]\unitlength1cm
\vspace*{-0.5cm}
\centerline{
%\put(-7.7,-0.3){$\text{(a)}$}
\epsfig{figure=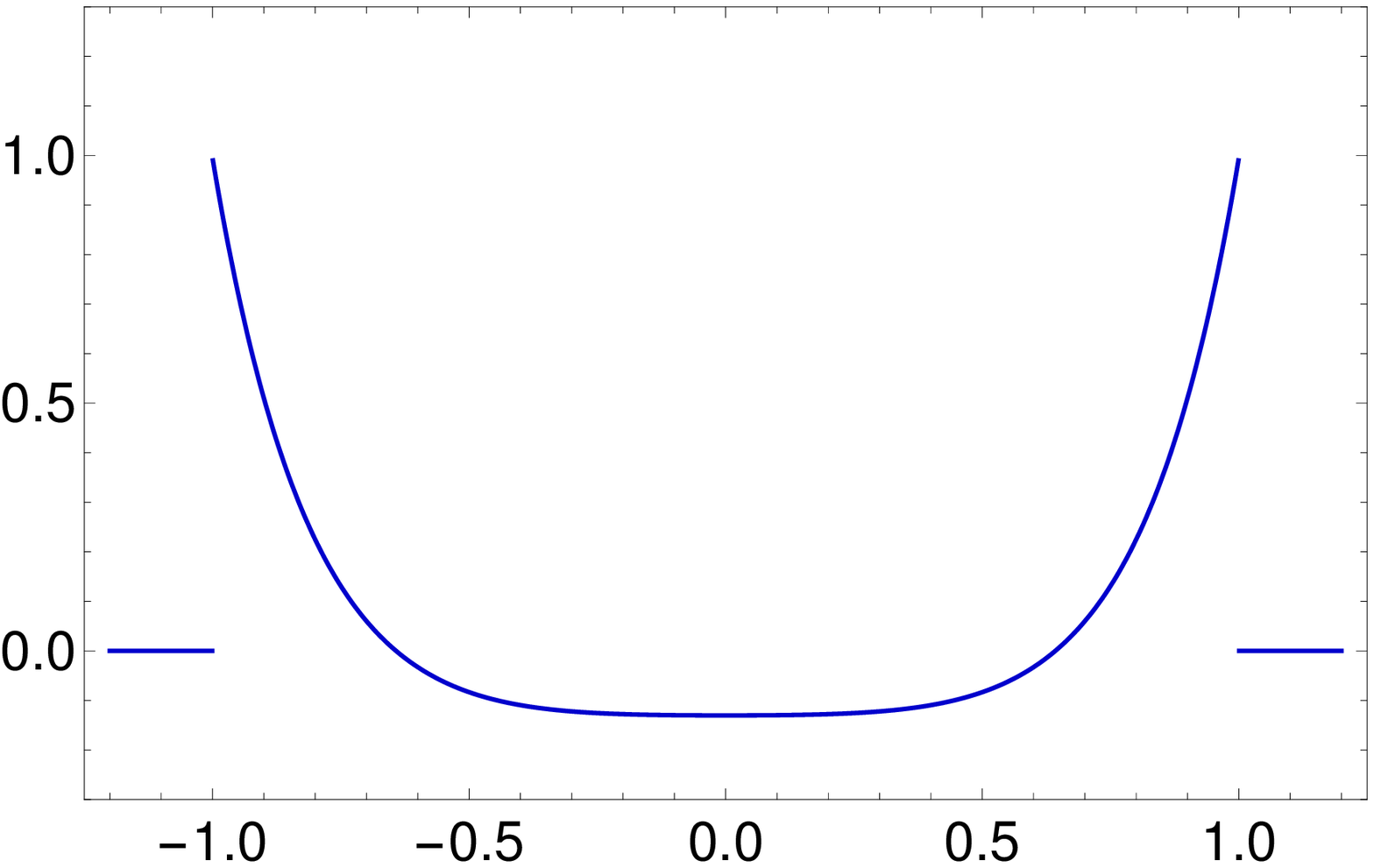,width=7.0cm}
%%\put(-1.0,1.0){$y/a$}
\put(-3.5,-0.4){$X$}
\put(-7.1,-0.3){$\text{(a)}$}
\hspace*{0.4cm}
\put(-0.1,-0.3){$\text{(b)}$}
%%\put(5.5,1.0){$y/a$}
%%\put(0.5,1.0){$x/a$}
\put(3.5,-0.4){$\tau$}
\epsfig{figure=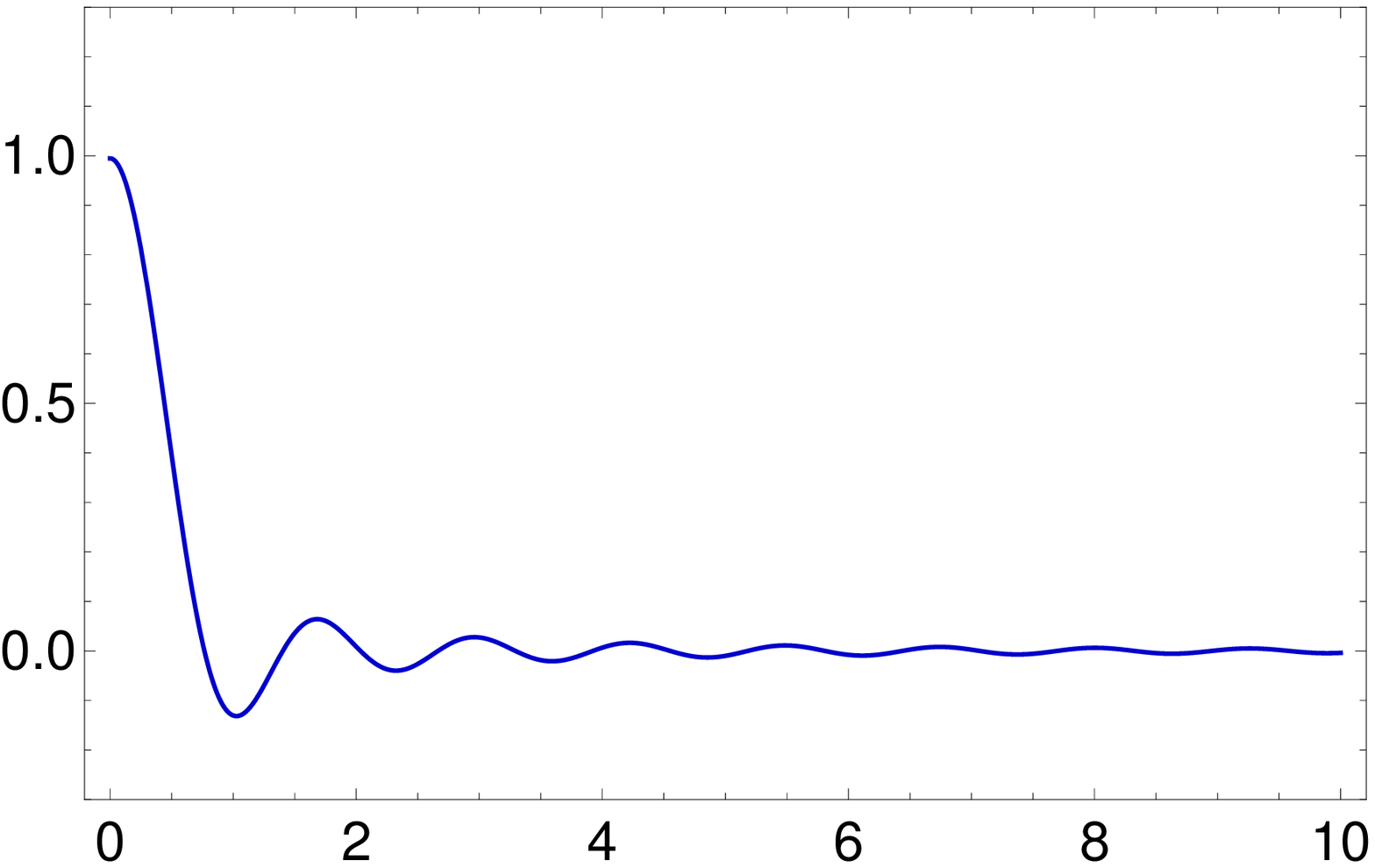,width=7.0cm}
%\put(0,-0.3){$\text{(b)}$}
}
\vspace*{0.2cm}
\centerline{
%\put(-7.0,-0.3){$\text{(c)}$}
\epsfig{figure=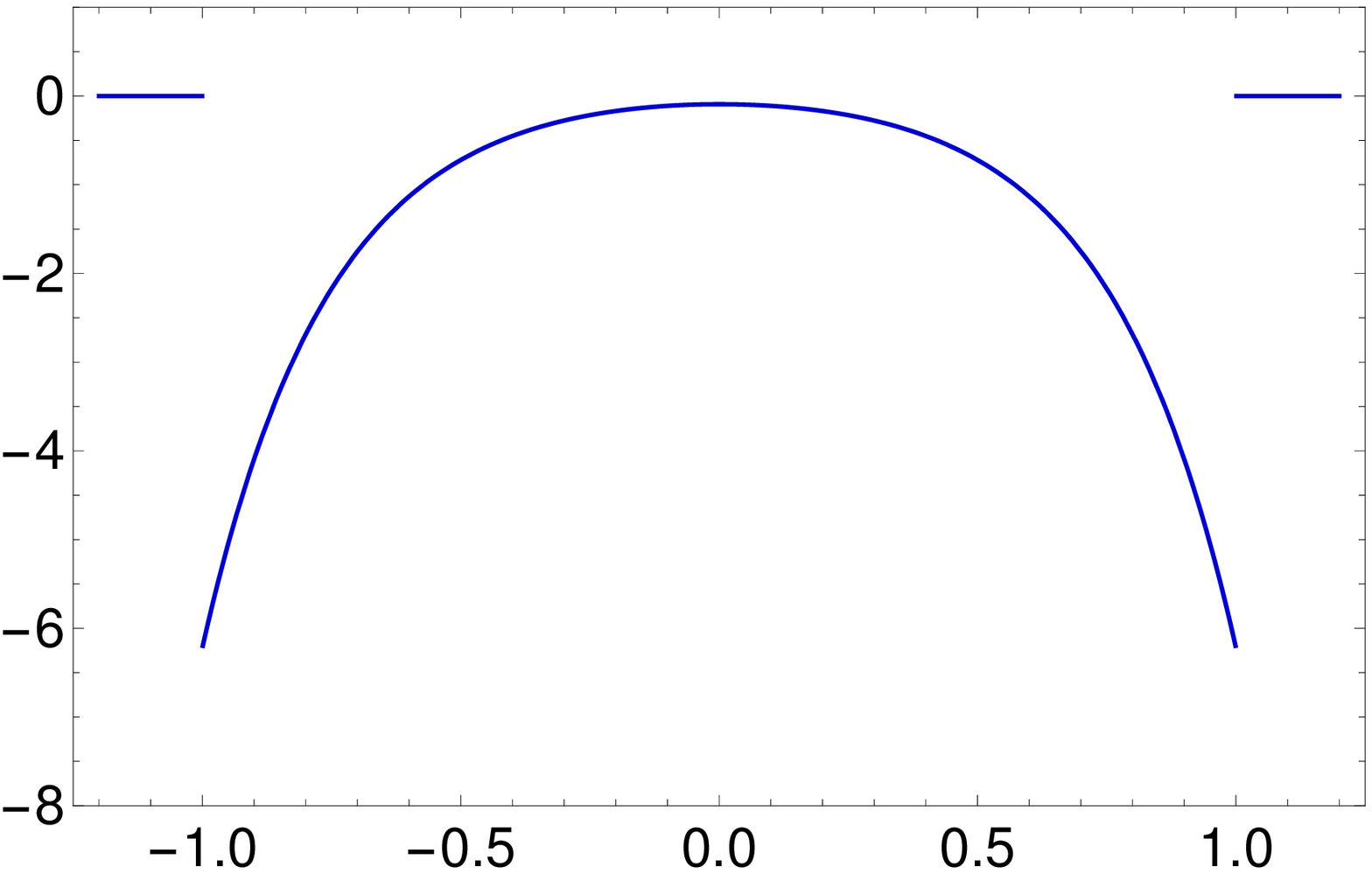,width=6.8cm}
%%\put(-1.0,1.0){$y/a$}
\put(-3.5,-0.4){$X$}
%%\put(-6.0,1.0){$x/a$}
\put(-7.1,-0.3){$\text{(c)}$}
\hspace*{0.4cm}
%%\put(5.5,1.0){$y/a$}
%%\put(0.5,1.0){$x/a$}
\put(3.5,-0.4){$\tau$}
\put(-0.1,-0.3){$\text{(d)}$}
\epsfig{figure=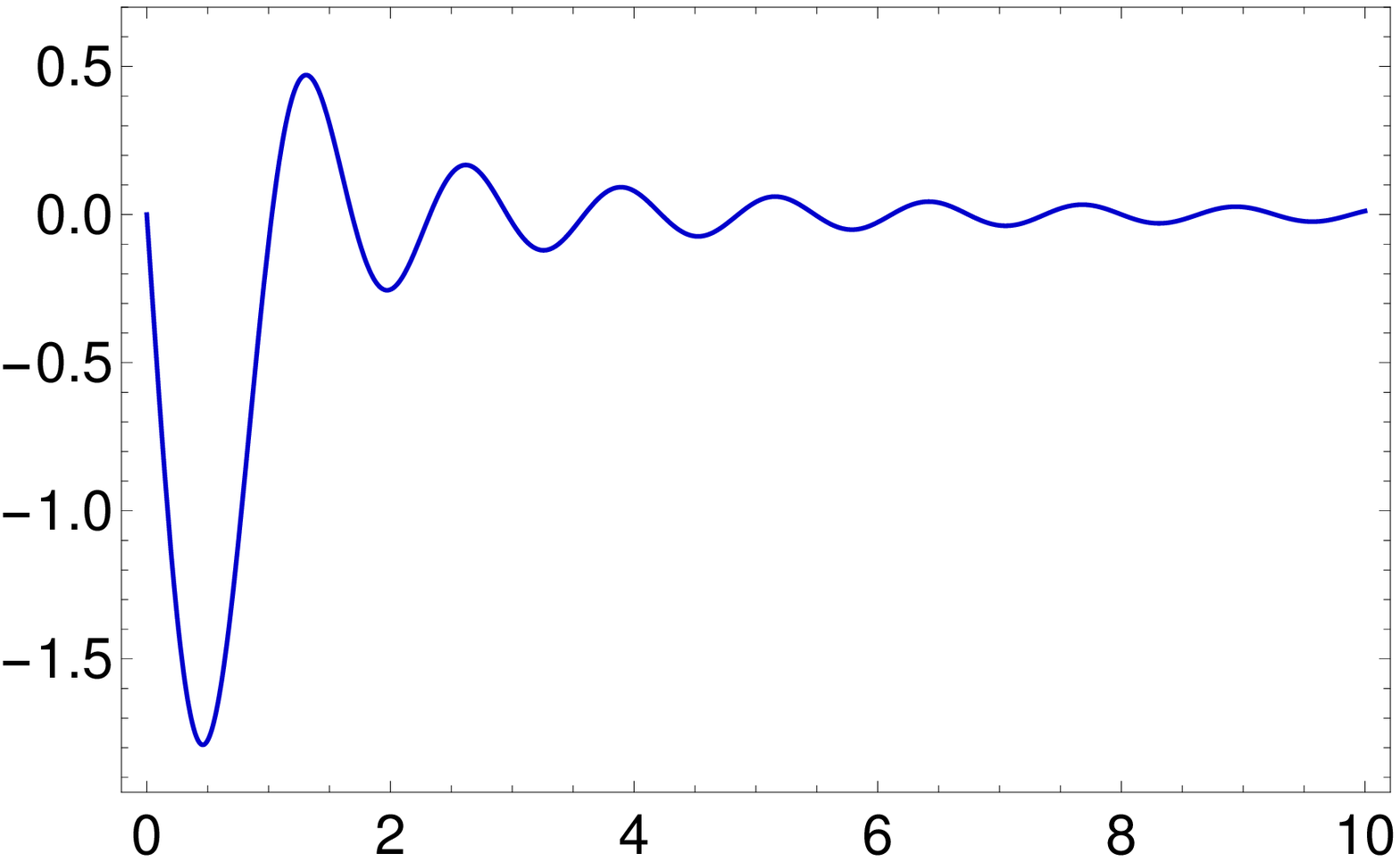,width=7.0cm}
}
\caption{Plots of the three-dimensional Bopp-Podolsky Green function 
for $c=1$, $\ell=0.2$: 
(a) $G^{\text{BP}}_{(3)}(\bm R,\tau=1)$ for $Y=Z=0$,
(b) $G^{\text{BP}}_{(3)}(\bm R=0,\tau)$,
(c) regular part of $\pd_\tau G^{\text{BP}}_{(3)}(\bm R,\tau=1)$
for $Y=Z=0$,
(d) regular part of $\pd_\tau G^{\text{BP}}_{(3)}(\bm R=0,\tau)$.}
%%(red dashed curves are the classical Green function $G_{(3)}^\square$).}
\label{fig:3D}
\end{figure}

Using 
\begin{align}
\label{rel-J1}
\lim_{z \to 0} \,\frac{1}{z}\,J_1(z)=\frac{1}{2}\,,
\end{align}
on the light cone, $c\tau=R$, 
the Green function~\eqref{G-BP-3d} is discontinuous (see Fig.~\ref{fig:3D}a) 
and reads as
\begin{align}
\label{G3-BP-LC}
G_{(3)}^{\rm BP}(\bm R, \tau)
\simeq\frac{c}{8\pi\ell^2}\, H\big(c\tau-R)\,.
\end{align}
Furthermore, the Green function~\eqref{G-BP-3d} 
shows a decreasing oscillation (see Fig.~\ref{fig:3D}b) and does not have a $\delta$-singularity 
unlike the Green function~\eqref{G-w-3d}.
One can say, Eq.~\eqref{G-BP-3d}  describes a wake in a plasma-like vacuum.

\subsection{2D Green functions}

\begin{figure}[t]\unitlength1cm
\vspace*{-0.5cm}
\centerline{
%\put(-7.7,-0.3){$\text{(a)}$}
\epsfig{figure=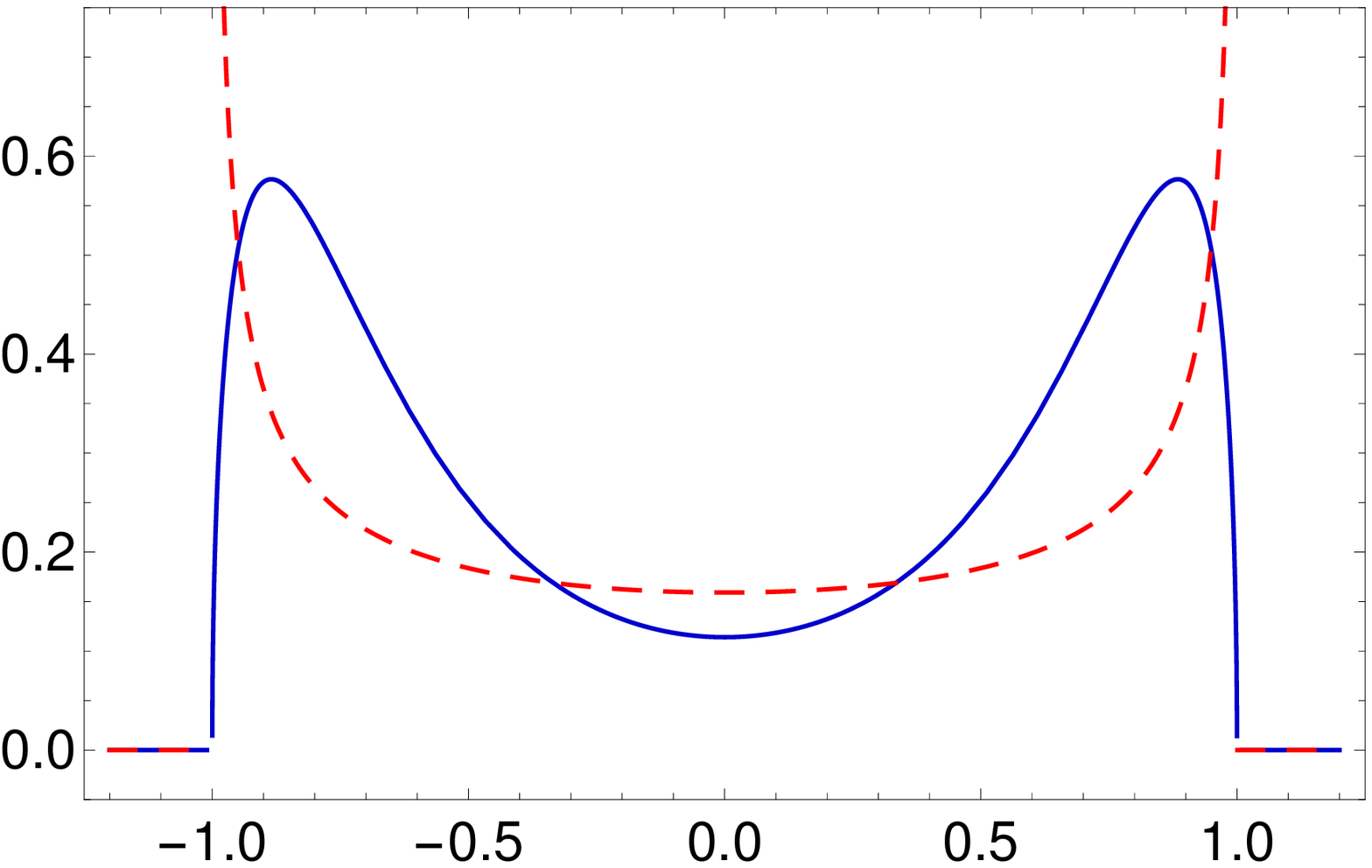,width=7.0cm}
%%\put(-1.0,1.0){$y/a$}
\put(-3.5,-0.4){$X$}
\put(-7.1,-0.3){$\text{(a)}$}
\hspace*{0.4cm}
\put(-0.1,-0.3){$\text{(b)}$}
%%\put(5.5,1.0){$y/a$}
%%\put(0.5,1.0){$x/a$}
\put(3.5,-0.4){$\tau$}
\epsfig{figure=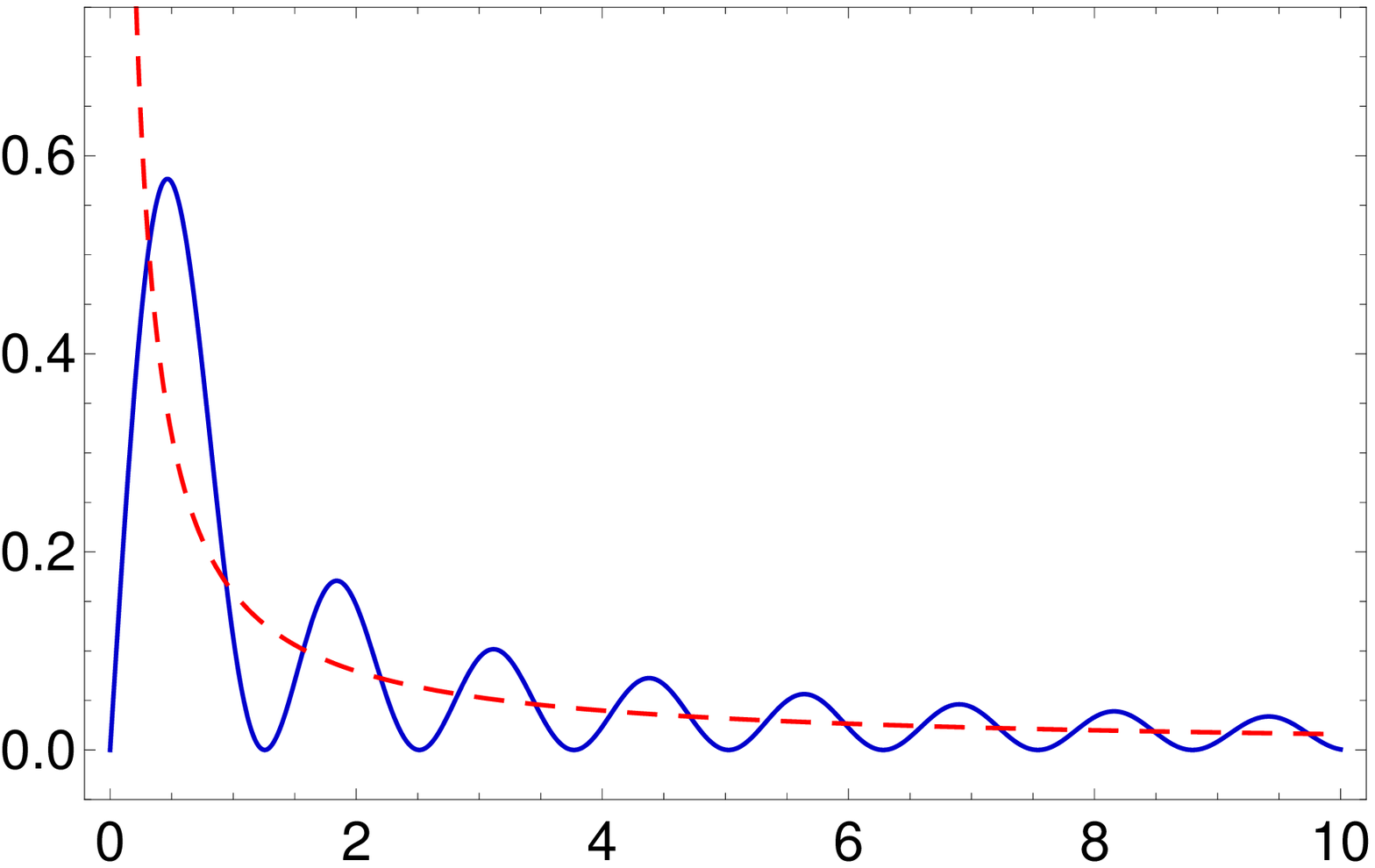,width=7.0cm}
%\put(0,-0.3){$\text{(b)}$}
}
\vspace*{0.2cm}
\centerline{
%\put(-7.0,-0.3){$\text{(c)}$}
\epsfig{figure=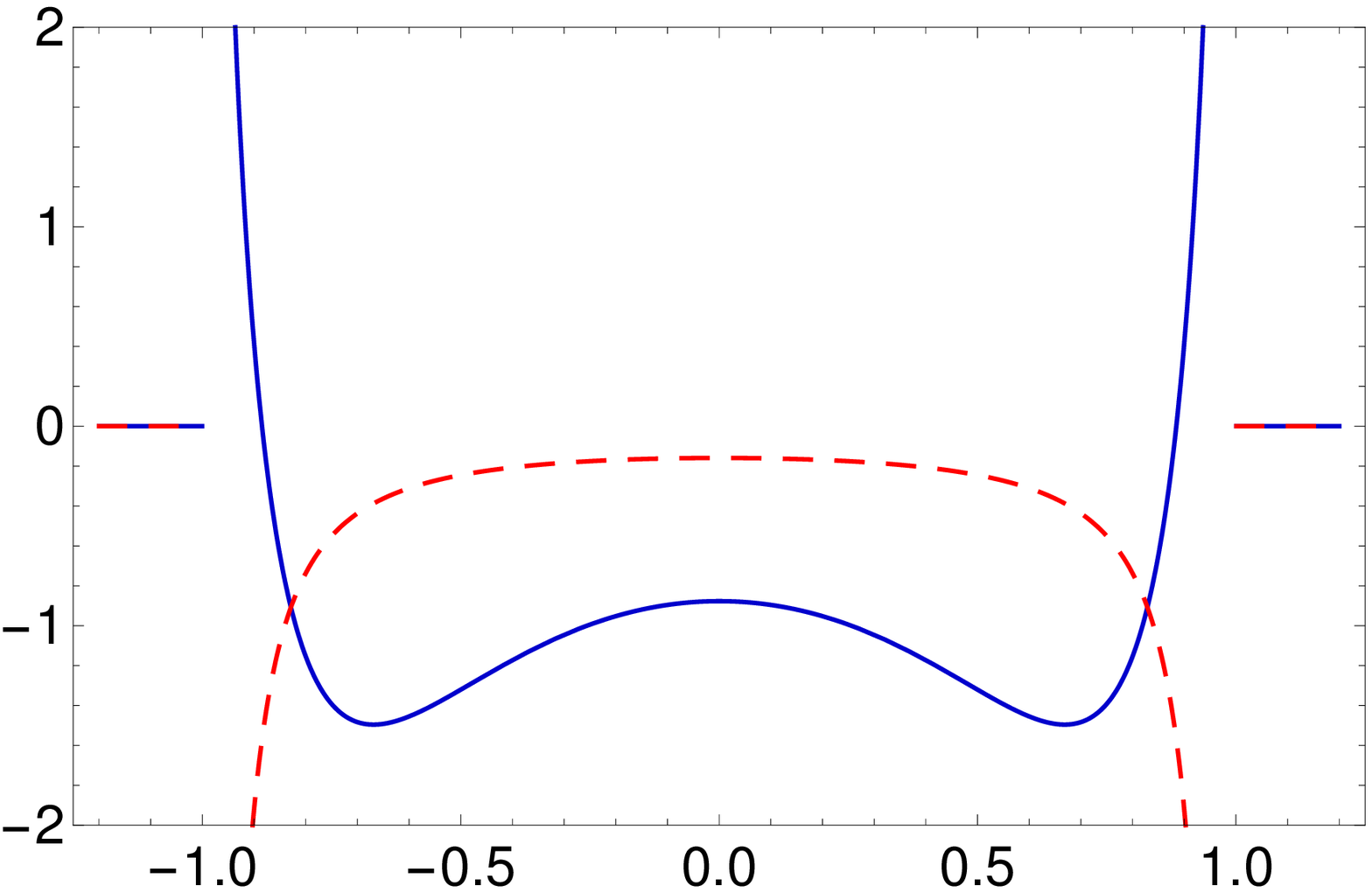,width=6.8cm}
%%\put(-1.0,1.0){$y/a$}
\put(-3.5,-0.4){$X$}
%%\put(-6.0,1.0){$x/a$}
\put(-7.1,-0.3){$\text{(c)}$}
\hspace*{0.4cm}
%%\put(5.5,1.0){$y/a$}
%%\put(0.5,1.0){$x/a$}
\put(3.5,-0.4){$\tau$}
\put(-0.1,-0.3){$\text{(d)}$}
\epsfig{figure=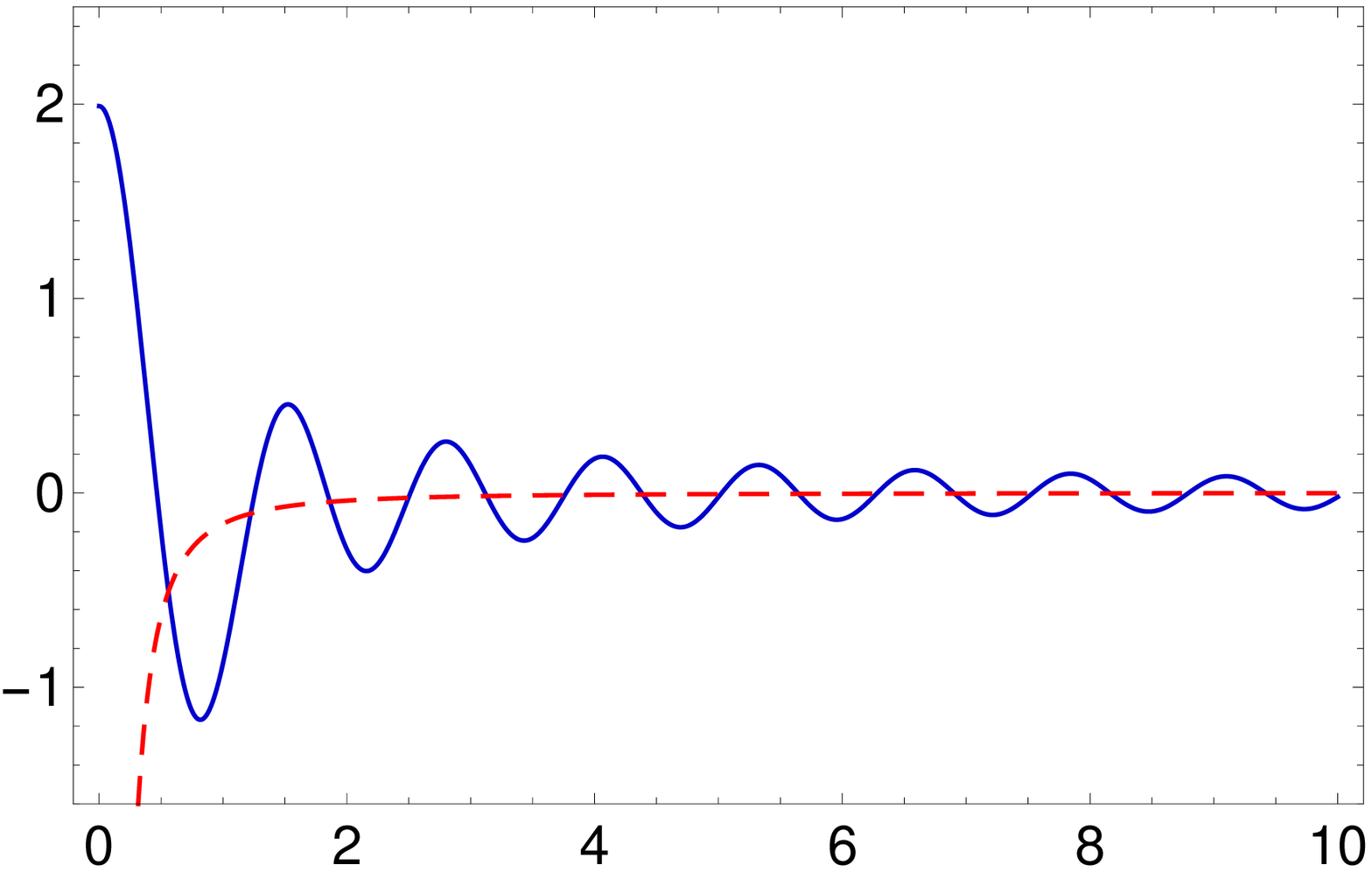,width=7.0cm}
}
\caption{Plots of the two-dimensional Bopp-Podolsky Green function 
for $c=1$, $\ell=0.2$: 
(a) $G^{\text{BP}}_{(2)}(\bm R,\tau=1)$ for $Y=0$,
(b) $G^{\text{BP}}_{(2)}(\bm R=0,\tau)$,
(c) $\pd_\tau G^{\text{BP}}_{(2)}(\bm R,\tau=1)$ for $Y=0$,
(d) $\pd_\tau G^{\text{BP}}_{(2)}(\bm R=0,\tau)$
(red dashed curves are the classical Green function $G_{(2)}^\square$
and  $\pd_\tau G_{(2)}^\square$).}
\label{fig:2D}
\end{figure}

The two-dimensional Green functions (fundamental solutions) of the 
wave (d'Alembert) operator~\eqref{wave}, the Klein-Gordon operator~\eqref{KGE}
and the Bopp-Podolsky (Klein-Gordon-d'Alembert) operator are the
(generalized) functions ($\tau>0$):
\begin{align}
\label{G-w-2d}
G_{(2)}^\square(\bm R, \tau)&=\frac{c}{2\pi}\,
\frac{H\big(c\tau-R\big)}{\sqrt{c^2\tau^2-R^2}}\,,\\
\label{G-KG-2d}
G_{(2)}^{\rm KG}(\bm R, \tau)
&=\frac{c}{2\pi\ell^2}\,
\frac{H\big(c\tau-R\big)}{\sqrt{c^2 \tau^2-R^2}}\, 
\cos\bigg( \frac{\sqrt{c^2 \tau^2-R^2}}{\ell}\bigg)\,,\\
%%\end{align}
%%where $H$ is the Heaviside step function.
\label{G-BP-2d}
G_{(2)}^{\rm BP}(\bm R, \tau)
&=\frac{c}{2\pi}\,
\frac{H\big(c\tau-R\big)}{\sqrt{c^2\tau^2-R^2}}\, 
\bigg[1-
\cos\bigg( \frac{\sqrt{c^2 \tau^2-R^2}}{\ell}\bigg)\bigg] \,,
\end{align}
where $R=\sqrt{(x-x')^2+(y-y')^2}$.
Eq.~\eqref{G-BP-2d} is obtained from Eq.~\eqref{BPG} using the Green functions~\eqref{G-w-2d} and \eqref{G-KG-2d}.
The Green function~\eqref{G-BP-2d} has been derived by~\citet{Lazar2010} in the framework of dislocation gauge theory.
The Green function~\eqref{G-BP-2d} of the Bopp-Podolsky operator is zero on
the light cone (see Fig.~\ref{fig:2D}a), since
\begin{align}
\lim_{z \to 0} \,\frac{1}{z}\,\cos(z)=\frac{1}{z}\,.
\end{align}
Furthermore, the Green function~\eqref{G-BP-2d} 
shows a decreasing oscillation around 
the classical Green function~\eqref{G-w-2d} (see Fig.~\ref{fig:2D}b).

\subsection{1D Green functions}

\begin{figure}[t]\unitlength1cm
\vspace*{-0.5cm}
\centerline{
%\put(-7.7,-0.3){$\text{(a)}$}
\epsfig{figure=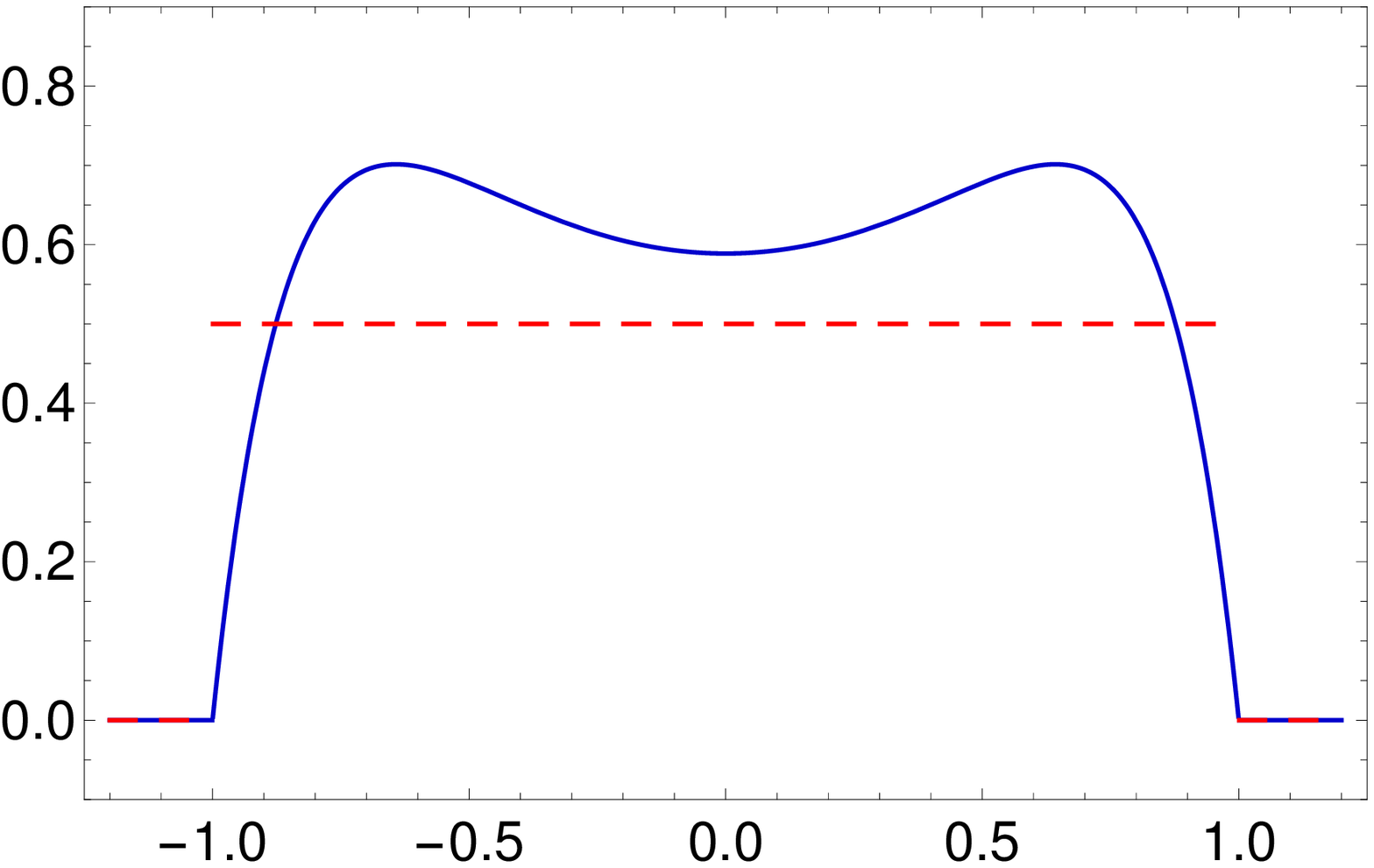,width=7.0cm}
%%\put(-1.0,1.0){$y/a$}
\put(-3.5,-0.4){$X$}
\put(-7.1,-0.3){$\text{(a)}$}
\hspace*{0.4cm}
\put(-0.1,-0.3){$\text{(b)}$}
%%\put(5.5,1.0){$y/a$}
%%\put(0.5,1.0){$x/a$}
\put(3.5,-0.4){$\tau$}
\epsfig{figure=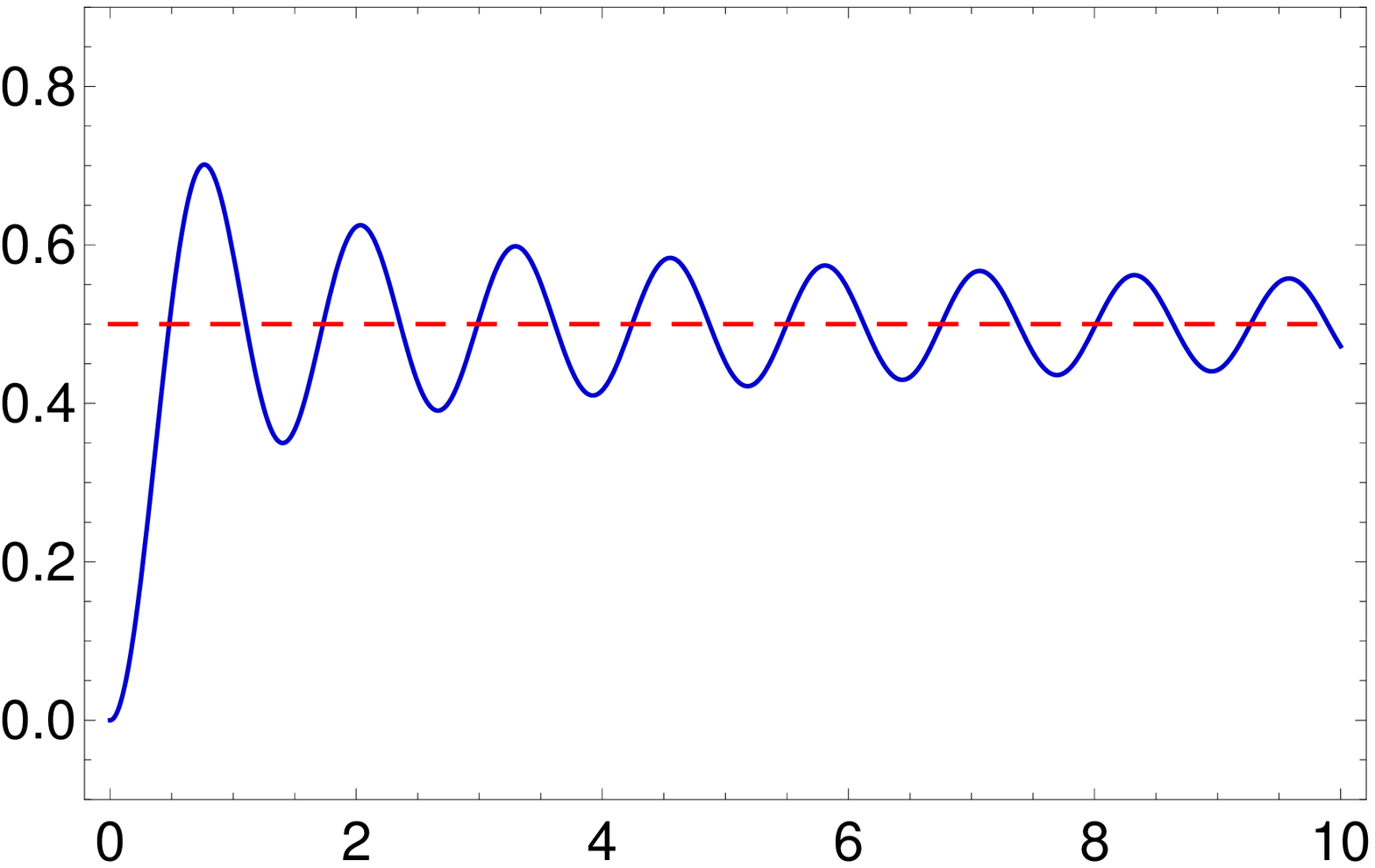,width=7.0cm}
%\put(0,-0.3){$\text{(b)}$}
}
\vspace*{0.2cm}
\centerline{
%\put(-7.0,-0.3){$\text{(c)}$}
\epsfig{figure=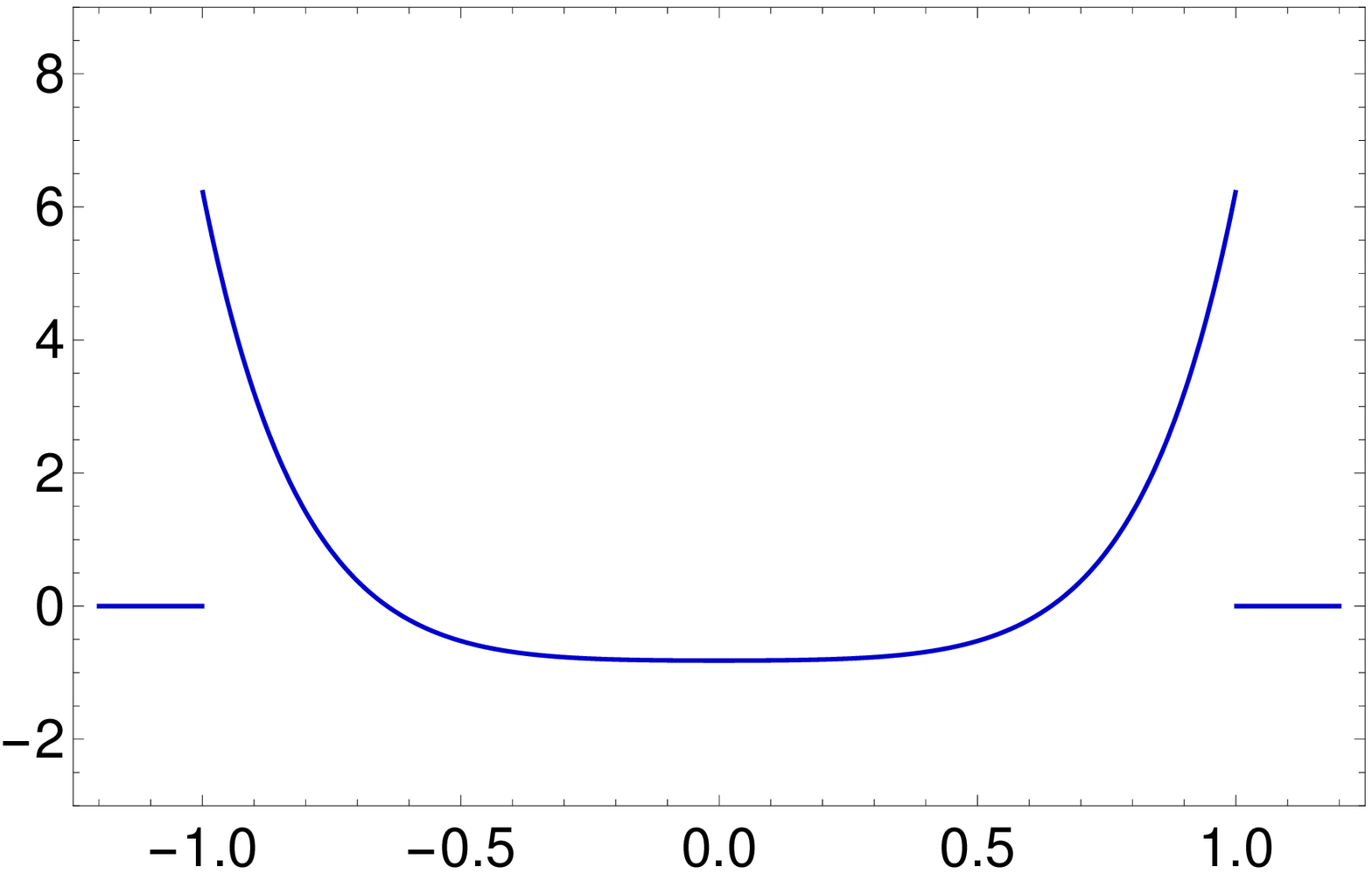,width=6.8cm}
%%\put(-1.0,1.0){$y/a$}
\put(-3.5,-0.4){$X$}
%%\put(-6.0,1.0){$x/a$}
\put(-7.1,-0.3){$\text{(c)}$}
\hspace*{0.4cm}
%%\put(5.5,1.0){$y/a$}
%%\put(0.5,1.0){$x/a$}
\put(3.5,-0.4){$\tau$}
\put(-0.1,-0.3){$\text{(d)}$}
\epsfig{figure=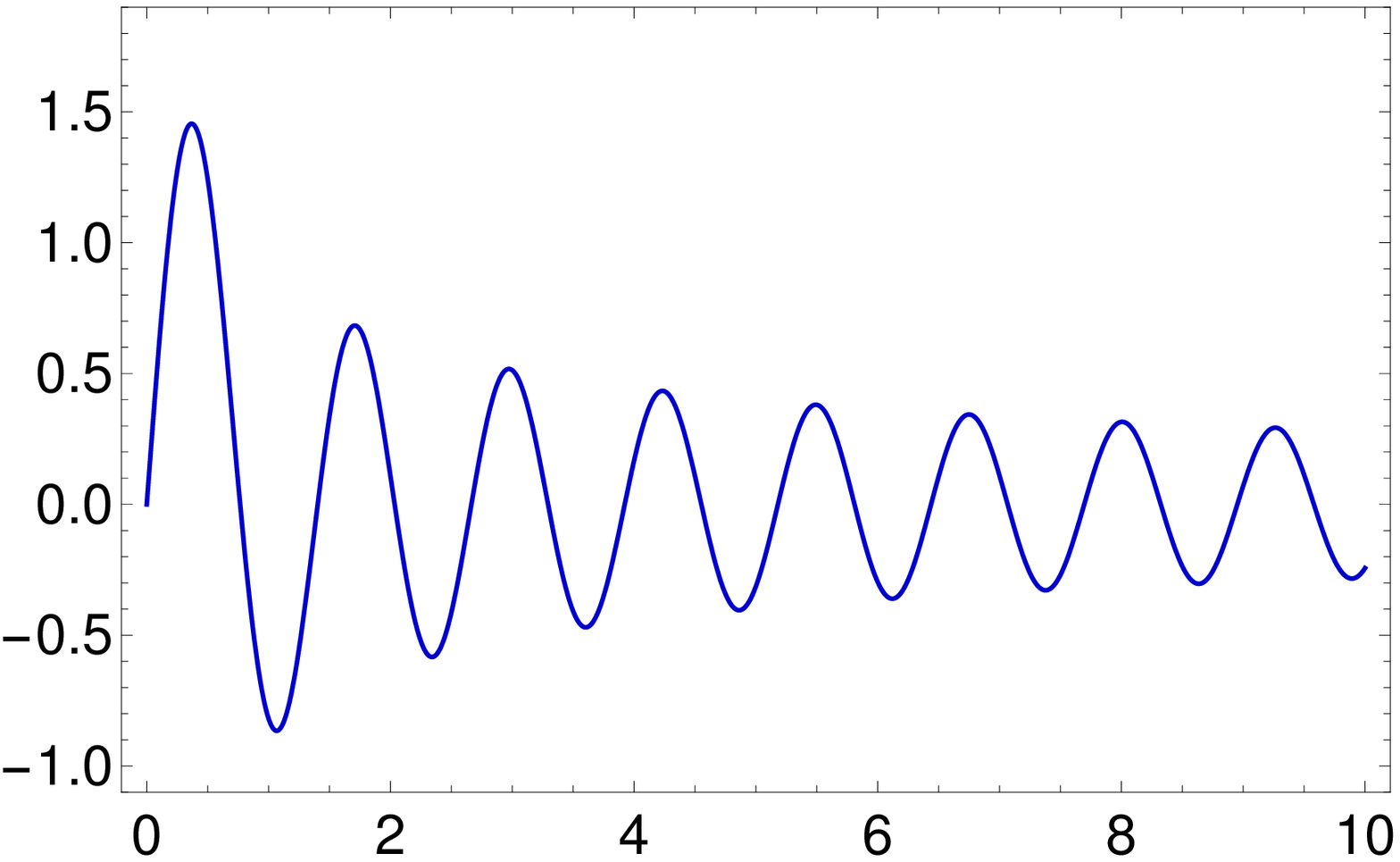,width=7.0cm}
}
\caption{Plots of the one-dimensional Bopp-Podolsky Green function 
for $c=1$, $\ell=0.2$: 
(a) $G^{\text{BP}}_{(1)}(X,\tau=1)$,
(b) $G^{\text{BP}}_{(1)}(X=0,\tau)$,
(c) $\pd_\tau G^{\text{BP}}_{(1)}(X,\tau=1)$,
(d) $\pd_\tau G^{\text{BP}}_{(1)}(X=0,\tau)$
(red dashed curves are the classical Green function $G_{(1)}^\square$).}
\label{fig:1D}
\end{figure}

The one-dimensional Green functions (fundamental solutions) of the 
wave (d'Alembert) operator~\eqref{wave}, the Klein-Gordon operator~\eqref{KGE}
and the Bopp-Podolsky (Klein-Gordon-d'Alembert) operator are the
(generalized) functions ($\tau>0$):
\begin{align}
\label{G-w-1d}
G_{(1)}^\square(X,\tau)&=\frac{c}{2}\, H\big(c\tau-|X|)\,,\\
\label{G-KG-1d}
G_{(1)}^{\rm KG}(X,\tau)
&=\frac{c}{2 \ell^2}\,
H\big(c\tau-|X|\big) \,
J_0 \bigg( \frac{\sqrt{c^2 \tau^2-X^2}}{\ell}\bigg)\,,\\
%%\end{align}
%%where $H$ is the Heaviside step function.
\label{G-BP-1d}
G_{(1)}^{\rm BP}(X,\tau)
&=\frac{c}{2}\, H\big(c\tau-|X|)\,
\bigg[1-J_0 \bigg( \frac{\sqrt{c^2 \tau^2-X^2}}{\ell}\bigg)\bigg]\,,
\end{align}
where $X=x-x'$ and $J_0$ is the Bessel function 
of the first kind of order zero. 
Eq.~\eqref{G-BP-1d} is obtained from Eq.~\eqref{BPG} using the Green functions~\eqref{G-w-1d} and \eqref{G-KG-1d}.
The Green function~\eqref{G-BP-1d} of the Bopp-Podolsky operator approaches zero on
the light cone (see Fig.~\ref{fig:1D}a), since
\begin{align}
\label{rel-J0}
\lim_{z \to 0} J_0(z)=1\,.
\end{align}
Due to the Bessel function term $J_0$, the Green function~\eqref{G-BP-1d} 
shows a decreasing oscillation around 
the classical Green function~\eqref{G-w-1d} (see Fig.~\ref{fig:1D}b).

\subsection{Derivatives of the Bopp-Podolsky Green function}
In this subsection, we derive the first time-derivative and first gradient of
the Bopp-Podolsky Green function.

\subsubsection{3D} 
The first time-derivative and first gradient
of the three-dimensional Bopp-Podolsky Green function~\eqref{G-BP-3d} read for $\tau>0$:
\begin{align}
\label{G-BP-3d-t}
\pd_\tau G_{(3)}^{\rm BP}(\bm R, \tau)
&=\frac{c^3 \tau}{4\pi\ell^2}\,\bigg[\frac{1}{2c\tau}\, \delta\big(c\tau-R\big)
-\frac{H\big(c\tau-R\big)}{(c^2 \tau^2-R^2)}\, 
J_2 \bigg(\frac{\sqrt{c^2 \tau^2-R^2}}{\ell}\bigg)\bigg]\,,\\
\label{G-BP-3d-grad}
\nabla G_{(3)}^{\rm BP}(\bm R, \tau)
&=-\frac{c \bm R}{4\pi\ell^2}\,\bigg[\frac{1}{2 R}\, \delta\big(c\tau-R\big)
- \frac{H\big(c\tau-R\big)}{(c^2 \tau^2-R^2)}\, 
J_2 \bigg(\frac{\sqrt{c^2 \tau^2-R^2}}{\ell}\bigg)\bigg]\,,
\end{align}
using $H'(z)=\delta(z)$, $\delta(z)f(z)=\delta(z)f(0)$, and 
$(J_1(z)/z)'=-J_2(z)/z$.  
$J_2$ is the Bessel function of the first kind of order two. 
Thus,  Eqs.~\eqref{G-BP-3d-t} and \eqref{G-BP-3d-grad} consist of two terms,
namely a Dirac $\delta$-term on the light cone plus a Bessel function term
inside the light cone.
The second parts (regular parts) of Eqs.~\eqref{G-BP-3d-t} and
\eqref{G-BP-3d-grad} are discontinuous and show a decreasing oscillation.

On the light cone, the derivatives of the Green function~$G_{(3)}^{\rm BP}$ 
possess a singularity of Dirac  $\delta$-type.  This is exhibited by the first term in Eqs.~\eqref{G-BP-3d-t} and \eqref{G-BP-3d-grad}.
The second term in Eqs.~\eqref{G-BP-3d-t} and \eqref{G-BP-3d-grad} 
is discontinuous on the light cone (see Fig.~\ref{fig:3D}c), since
\begin{align}
\label{J2-rel}
\lim_{z \to 0} \,\frac{1}{z^2}\,J_2(z)=\frac{1}{8}\,.
\end{align}
%%On the light cone, 
In the neighbourhood of the light cone,
Eqs.~\eqref{G-BP-3d-t} and \eqref{G-BP-3d-grad} 
have the form 
\begin{align}
\label{G-BP-3d-t-LC}
\pd_\tau G_{(3)}^{\rm BP}(\bm R, \tau)
&\simeq\frac{c^3 \tau}{4\pi\ell^2}\,\bigg[\frac{1}{2c\tau}\, \delta\big(c\tau-R\big)
-\frac{1}{8\ell^2}\, H\big(c\tau-R\big)\bigg]\,,\\
\label{G-BP-3d-grad-LC}
\nabla G_{(3)}^{\rm BP}(\bm R, \tau)
&\simeq-\frac{c \bm R}{4\pi\ell^2}\,\bigg[\frac{1}{2 R}\, \delta\big(c\tau-R\big)
-\frac{1}{8\ell^2}\, H\big(c\tau-R\big)\bigg]\,.
\end{align}
It can be seen in Fig.~\ref{fig:3D}d that
the second parts (regular parts) of Eqs.~\eqref{G-BP-3d-t} and
\eqref{G-BP-3d-grad} show a decreasing oscillation.

\subsubsection{2D} 
The first time-derivative and first gradient
of the two-dimensional Bopp-Podolsky Green function~\eqref{G-BP-2d} read for $\tau>0$:
\begin{align}
\label{G-BP-2d-t}
\pd_\tau G_{(2)}^{\rm BP}(\bm R, \tau)
&=-\frac{c^3\tau}{2\pi}\,
\frac{H\big(c\tau-R\big)}{(c^2\tau^2-R^2)^{\frac{3}{2}}}\, 
\bigg[
1-\cos\bigg( \frac{\sqrt{c^2 \tau^2-R^2}}{\ell}\bigg)
-\frac{\sqrt{c^2\tau^2-R^2}}{\ell}
\, \sin\bigg( \frac{\sqrt{c^2 \tau^2-R^2}}{\ell}\bigg)
\bigg] \,,\\
%%\end{align}
%%\begin{align}
\label{G-BP-2d-grad}
\nabla G_{(2)}^{\rm BP}(\bm R, \tau)
&=\frac{c\bm R}{2\pi}\,
\frac{H\big(c\tau-R\big)}{(c^2\tau^2-R^2)^{\frac{3}{2}}}\, 
\bigg[
1-\cos\bigg( \frac{\sqrt{c^2 \tau^2-R^2}}{\ell}\bigg)
-\frac{\sqrt{c^2\tau^2-R^2}}{\ell}
\, \sin\bigg( \frac{\sqrt{c^2 \tau^2-R^2}}{\ell}\bigg)
\bigg] \,.
\end{align}

On the light cone, the derivatives of the Green function~$G_{(2)}^{\rm BP}$ 
possess a $1/z$-singularity  (see Fig.~\ref{fig:2D}c), since
\begin{align}
\lim_{z \to 0} \,\frac{1}{z^3}\,\cos(z)=\frac{1}{z^3}-\frac{1}{2z}
\end{align}
and
\begin{align}
\lim_{z \to 0} \,\frac{1}{z^2}\,\sin(z)=\frac{1}{z}\,,
\end{align}
they are discontinuous.
Of course, the $1/z$-singularity is weaker than the non-integrable
$1/z^3$-singularity.
%%On the light cone, 
In the neighbourhood of the light cone,
 Eqs.~\eqref{G-BP-2d-t} and \eqref{G-BP-2d-grad} have the form 
\begin{align}
\label{G-BP-2d-t-LC}
\pd_\tau G_{(2)}^{\rm BP}(\bm R, \tau)
&\simeq\frac{c^3\tau}{4\pi\ell^2}\,
\frac{H\big(c\tau-R\big)}{\sqrt{c^2\tau^2-R^2}}\,,\\
\label{G-BP-2d-grad-LC}
\nabla G_{(2)}^{\rm BP}(\bm R, \tau)
&\simeq-\frac{c\bm R}{4\pi\ell^2}\,
\frac{H\big(c\tau-R\big)}{\sqrt{c^2\tau^2-R^2}}\,.
\end{align}
Furthermore,  Eqs.~\eqref{G-BP-2d-t} and \eqref{G-BP-2d-grad}
show a decreasing oscillation around 
the classical singularity (see Fig.~\ref{fig:2D}d).

\subsubsection{1D}

The first time-derivative and space-derivative
of the one-dimensional Bopp-Podolsky Green function~\eqref{G-BP-1d} read for $\tau>0$:  
\begin{align}
\label{G-BP-1d-t}
\pd_\tau G_{(1)}^{\rm BP}(X,\tau)
&=\frac{c^3 \tau}{2\ell}\, 
\frac{H\big(c\tau-|X|)}{\sqrt{c^2\tau^2-X^2}}\,
J_1 \bigg( \frac{\sqrt{c^2 \tau^2-X^2}}{\ell}\bigg)\,,\\
%%\end{align}
%%\begin{align}
\label{G-BP-1d-grad}
\pd_X G_{(1)}^{\rm BP}(X,\tau)
&=-\frac{c X}{2\ell}\, 
\frac{H\big(c\tau-|X|)}{\sqrt{c^2\tau^2-X^2}}\,
J_1 \bigg( \frac{\sqrt{c^2 \tau^2-X^2}}{\ell}\bigg)\,,
\end{align}
using $J_0'=-J_1$.

On the light cone, the derivatives of the Green function~$G_{(1)}^{\rm BP}$,
Eqs.~\eqref{G-BP-1d-t} and \eqref{G-BP-1d-grad}, 
have a jump discontinuity, due to Eq.~\eqref{rel-J1}
(see Fig.~\ref{fig:1D}c), namely 
\begin{align}
\label{G-BP-1d-t-LC}
\pd_\tau G_{(1)}^{\rm BP}(X,\tau)
&\simeq\frac{c^3 \tau}{4\ell^2}\, H\big(c\tau-|X|)\,,\\
\label{G-BP-1d-grad-LC}
\pd_X G_{(1)}^{\rm BP}(X,\tau)
&\simeq-\frac{c X}{4\ell^2}\, H\big(c\tau-|X|)\,.
\end{align}
Furthermore, Eqs.~\eqref{G-BP-1d-t} and \eqref{G-BP-1d-grad}
show a decreasing oscillation (see Fig.~\ref{fig:1D}d) unlike the derivative
of the Green function of the d'Alembert equation given in terms of $\delta\big(c\tau-|X|)$.

\section{Retarded potentials and retarded electromagnetic field strengths}
\label{sec4}

Solutions based on retarded Green functions lead to retarded fields (like retarded potentials and retarded electromagnetic field strengths)
in the form of retarded integrals.
Retarded integrals are mathematical expressions reflecting the phenomenon of
``finite signal speed" (e.g. \citep{Jefimenko}).

\subsection{Retarded potentials}

The retarded electromagnetic potentials are the solutions of the inhomogeneous Bopp-Podolsky equations ~\eqref{phi-w} and \eqref{A-w} 
and for zero initial conditions they are given as convolution of the
(retarded) Green function $G^{\rm BP}$ and 
the given charge and current densities ($\rho$, $\bm J$) 
\begin{align}
\label{phi-rp}
\phi&=\frac{1}{\varepsilon_0}\, G^{\rm BP}*\rho\,,\\
\label{A-rp}
\bm A&=\mu_0\, G^{\rm BP}*\bm J\,.
\end{align}
%%%where $*$ denotes the convolution in space and time.
Explicitly, the convolution integrals~\eqref{phi-rp} and \eqref{A-rp} read as 
\begin{align}
\label{phi-rp-2}
\phi_{(n)}(\bm r,t)&=\frac{1}{\varepsilon_0}\, \int_{-\infty}^t\d t'\int_{\Bbb R^n} \d \bm r'\, G_{(n)}^{\rm BP}(\bm r-\bm r',t-t')\,\rho(\bm r',t')\,,\\
\label{A-rp-2}
\bm A_{(n)}(\bm r,t)&=\mu_0 \int_{-\infty}^t\d t'\int_{\Bbb R^n} \d \bm r'\, G_{(n)}^{\rm BP}(\bm r-\bm r',t-t') \,\bm J(\bm r',t')\,,
\end{align}
where $\bm r'$ is the source point and $\bm r$ is the field point. Here $n$ denotes the spatial dimension.

Substituting Eqs.~\eqref{phi-rp} and \eqref{A-rp}
into the generalized Lorentz gauge condition~\eqref{LG} and using
Eqs.~\eqref{BPE-2} and \eqref{CE}, it can be seen that the 
generalized Lorentz gauge condition is satisfied
\begin{align}
\label{LG-2}
\big[1+\ell^2 \square \big]
\left(\frac{1}{c^2}\,\pd_t \phi+ \nabla \cdot \bm A\right)
&=\mu_0 \big[1+\ell^2 \square \big] G^{\rm BP}
*\big(\pd_t \rho+ \nabla \cdot \bm J\big)\nonumber\\
&=\mu_0 \big(\pd_t \rho+ \nabla \cdot \bm J\big)*G^{\square}
\nonumber\\
&=0\,.
\end{align}

\subsubsection{3D} 

Substituting the Bopp-Podolsky Green function~\eqref{G-BP-3d} into Eqs.~\eqref{phi-rp-2} and \eqref{A-rp-2}, 
the three-dimensional retarded electromagnetic potentials read as 
\begin{align}
\label{phi-ret-3d}
\phi_{(3)}(\bm r,t)
&=\frac{c}{4\pi\varepsilon_0\ell}\int_{-\infty}^t\d t'\int_{\Bbb R^3} \d \bm r'\,
\frac{H\big(c\tau-R\big)}{\sqrt{c^2 \tau^2-R^2}}\, 
J_1 \bigg(\frac{\sqrt{c^2 \tau^2-R^2}}{\ell}\bigg) \rho(\bm r',t')\nonumber\\
&=\frac{c}{4\pi\varepsilon_0\ell}\int_{-\infty}^{t-R/c}\d t'\int_{\Bbb R^3} \d \bm r'\,
\frac{\rho(\bm r',t')}{\sqrt{c^2 \tau^2-R^2}}\, 
J_1 \bigg(\frac{\sqrt{c^2 \tau^2-R^2}}{\ell}\bigg) 
\end{align}
and
\begin{align}
\label{A-ret-3d}
\bm A_{(3)}(\bm r,t)
&=\frac{\mu_0 c}{4\pi\ell}\int_{-\infty}^t\d t'\int_{\Bbb R^3} \d \bm r'\,
\frac{H\big(c\tau-R\big)}{\sqrt{c^2 \tau^2-R^2}}\, 
J_1 \bigg(\frac{\sqrt{c^2 \tau^2-R^2}}{\ell}\bigg) \bm J(\bm r',t')\nonumber\\
&=\frac{\mu_0 c}{4\pi\ell}\int_{-\infty}^{t-R/c}\d t'\int_{\Bbb R^3} \d \bm r'\,
\frac{\bm J(\bm r',t')}{\sqrt{c^2 \tau^2-R^2}}\, 
J_1 \bigg(\frac{\sqrt{c^2 \tau^2-R^2}}{\ell}\bigg) 
\end{align}
since $H(c\tau -R)=0$ for $t'>t-R/c$.
In the Bopp-Podolsky electrodynamics,  the three-dimensional 
retarded potentials~\eqref{phi-ret-3d}
and \eqref{A-ret-3d} possess an afterglow,
since they draw contribution emitted at all times $t'$ from $-\infty$ up to $t-R/c$. 
The retarded time is a result of the finite speed of propagation for electromagnetic signals.

\subsubsection{2D} 

Substituting the Bopp-Podolsky Green function~\eqref{G-BP-2d} into Eqs.~\eqref{phi-rp-2} and \eqref{A-rp-2}, 
the two-dimensional retarded electromagnetic potentials become
\begin{align}
\label{phi-ret-2d}
\phi_{(2)}(\bm r,t)
&=\frac{c}{2\pi\varepsilon_0}\int_{-\infty}^t\d t'\int_{\Bbb R^2} \d \bm r'\,
\frac{H\big(c\tau-R\big)}{\sqrt{c^2\tau^2-R^2}}\, 
\bigg[1-\cos\bigg( \frac{\sqrt{c^2 \tau^2-R^2}}{\ell}\bigg)\bigg]\rho(\bm
r',t') 
\nonumber\\
&=\frac{c}{2\pi\varepsilon_0}\int_{-\infty}^{t-R/c}\d t'\int_{\Bbb R^2} \d \bm r'\,
\frac{\rho(\bm r',t')}{\sqrt{c^2\tau^2-R^2}}\, 
\bigg[1-\cos\bigg( \frac{\sqrt{c^2 \tau^2-R^2}}{\ell}\bigg)\bigg] 
\end{align}
and
\begin{align}
\label{A-ret-2d}
\bm A_{(2)}(\bm r,t)
&=\frac{\mu_0 c}{2\pi}\int_{-\infty}^t\d t'\int_{\Bbb R^2} \d \bm r'\,
\frac{H\big(c\tau-R\big)}{\sqrt{c^2\tau^2-R^2}}\, 
\bigg[1-\cos\bigg( \frac{\sqrt{c^2 \tau^2-R^2}}{\ell}\bigg)\bigg]\bm J(\bm r',t') \nonumber\\
&=\frac{\mu_0 c}{2\pi}\int_{-\infty}^{t-R/c}\d t'\int_{\Bbb R^2} \d \bm r'\,
\frac{\bm J(\bm r',t')}{\sqrt{c^2\tau^2-R^2}}\, 
\bigg[1-\cos\bigg( \frac{\sqrt{c^2 \tau^2-R^2}}{\ell}\bigg)\bigg]
\end{align}
since $H(c\tau -R)=0$ for $t'>t-R/c$.
%%In the Bopp-Podolsky electrodynamics, 
Thus, the two-dimensional 
retarded potentials~\eqref{phi-ret-2d}
and \eqref{A-ret-2d} show an afterglow,
since they draw contribution emitted at all times $t'$ from $-\infty$ up to $t-R/c$.

\subsubsection{1D}

In the version of the Bopp-Podolsky electrodynamics in one spatial dimension,  
the potentials $\phi_{(1)}$ and $A_{(1)}$ are both a scalar field, and the current
density $J$ is also a scalar field.

Substituting the Bopp-Podolsky Green function~\eqref{G-BP-1d} into Eqs.~\eqref{phi-rp-2} and \eqref{A-rp-2}, 
the one-dimensional retarded electromagnetic potentials read 
\begin{align}
\label{phi-ret-1d}
\phi_{(1)}(x,t)
&=\frac{c}{2\varepsilon_0}\int_{-\infty}^t\d t'\int_{-\infty}^\infty\d x'
\, H\big(c\tau-|X|)\,
\bigg[1-J_0 \bigg( \frac{\sqrt{c^2
      \tau^2-X^2}}{\ell}\bigg)\bigg]\rho(x',t')\,\nonumber\\
&=\frac{c}{2\varepsilon_0}\int_{-\infty}^{t-|X|/c} \d t'\int_{-\infty}^\infty\d x'
\,\rho(x',t')
\bigg[1-J_0 \bigg( \frac{\sqrt{c^2
      \tau^2-X^2}}{\ell}\bigg)\bigg]
\end{align}
and
\begin{align}
\label{A-ret-1d}
A_{(1)}(x,t)
&=\frac{\mu_0 c}{2}\int_{-\infty}^t\d t'\int_{-\infty}^\infty\d x'
\, H\big(c\tau-|X|)\,
\bigg[1-J_0 \bigg( \frac{\sqrt{c^2
      \tau^2-X^2}}{\ell}\bigg)\bigg] J(x',t')\,\nonumber\\
&=\frac{\mu_0 c}{2}\int_{-\infty}^{t-|X|/c}\d t'\int_{-\infty}^\infty\d x'
\, J(x',t')
\bigg[1-J_0 \bigg( \frac{\sqrt{c^2\tau^2-X^2}}{\ell}\bigg)\bigg] 
\end{align}
since $H(c\tau -|X|)=0$ for $t'>t-|X|/c$.
%%In the Bopp-Podolsky electrodynamics,  
It can be seen that the one-dimensional 
retarded potentials~\eqref{phi-ret-1d} and \eqref{A-ret-1d} draw contribution emitted at all times $t'$ from $-\infty$ up to $t-|X|/c$.

In the Bopp-Podolsky electrodynamics,  the retarded potentials possess an afterglow in 1D, 2D and 3D since they draw contribution emitted
at all times $t'$ from $-\infty$ up to $t-R/c$ 
unlike in the classical Maxwell electrodynamics where only 
the retarded potentials possess an afterglow in 1D and 2D (see, e.g., \citep{Barton,Wl}). 

\subsection{Retarded electromagnetic field strengths}

Substituting Eqs.~\eqref{phi-rp} and \eqref{A-rp} into 
the electromagnetic fields~\eqref{E} and \eqref{B} or solving 
Eqs.~\eqref{E-w} and \eqref{B-w}, the electromagnetic fields ($\bm E$, $\bm B$)
are given by the convolution of the Green function~$G^{\rm BP}$ and 
the given charge and current densities ($\rho$, $\bm J$) and read as 
\begin{align}
\label{E-rp}
\bm E_{(n)}(\bm r,t)&=-\frac{1}{\varepsilon_0}\, \int_{-\infty}^t\d t'\int_{\Bbb R^n}
\d \bm r'\, 
\Big(\nabla G_{(n)}^{\rm BP}(\bm r-\bm r',t-t')\,\rho(\bm r',t')
\nonumber\\
&\qquad\qquad\qquad\qquad
+\frac{1}{c^2} \pd_t  G_{(n)}^{\rm BP}(\bm r-\bm r',t-t')\, \bm J(\bm r',t')\Big)
\,,\\
\label{B-rp}
\bm B_{(n)}(\bm r,t)&=\mu_0 \int_{-\infty}^t\d t'\int_{\Bbb R^n} \d \bm r'\,
\nabla G_{(n)}^{\rm BP}(\bm r-\bm r',t-t') \times \bm J(\bm r',t')\,.
\end{align}

\subsubsection{3D}

Substituting the derivatives of the Bopp-Podolsky Green function~\eqref{G-BP-3d-t} 
and \eqref{G-BP-3d-grad} into Eqs.~\eqref{E-rp} and \eqref{B-rp}, 
the three-dimensional retarded electromagnetic field strengths read as 
\begin{align}
\label{E-ret-3d}
\bm E_{(3)}(\bm r,t)
&=\frac{1}{8\pi\varepsilon_0 \ell^2}\,
\int_{\Bbb R^3}\d \bm r'\, 
\bigg(\frac{\bm R}{R} \, \rho\big(\bm r',t-R/c\big)
-\frac{1}{c}\, \bm J\big(\bm r',t-R/c\big)\bigg)\nonumber\\
&\quad
-\frac{c}{4\pi\varepsilon_0 \ell^2}\,
\int_{-\infty}^{t-R/c}\d t'\int_{\Bbb R^3}\d \bm r'\, 
\frac{\big[\bm R \rho(\bm r',t')-\tau \bm J(\bm r',t')\big]}
{(c^2 \tau^2-R^2)}\, 
J_2 \bigg(\frac{\sqrt{c^2 \tau^2-R^2}}{\ell}\bigg)
\end{align}
and 
\begin{align}
\label{B-ret-3d}
\bm B_{(3)}(\bm r,t)
&=-\frac{\mu_0}{8\pi\ell^2}\,
\int_{\Bbb R^3}\d \bm r'\, 
\frac{\bm R}{R}\times \bm J\big(\bm r',t-R/c\big)\nonumber\\
&\quad 
+\frac{\mu_0 c}{4\pi\ell^2}\,
\int_{-\infty}^{t-R/c}\d t'\int_{\Bbb R^3}\d \bm r'\, 
\frac{\bm R \times \bm J(\bm r',t')}{(c^2 \tau^2-R^2)}\, 
J_2 \bigg(\frac{\sqrt{c^2 \tau^2-R^2}}{\ell}\bigg)
\,.
\end{align}
In the first part of Eqs.~\eqref{E-ret-3d} and \eqref{B-ret-3d}, 
the $\delta$-function in Eqs.~\eqref{G-BP-3d-t} and \eqref{G-BP-3d-grad}
picked out the value of $\rho$ and $\bm J$ at the retarded time,
$t-R/c$, which is earlier than $t$ 
by as long as it takes a signal with speed $c$ to travel 
from the source point $\rr'$ to the field point $\rr$.
From each point $\rr'$, the first part of Eqs.~\eqref{E-ret-3d} and \eqref{B-ret-3d} 
draws contributions emitted at the retarded time $t-R/c$.
The second part of Eqs.~\eqref{E-ret-3d} and \eqref{B-ret-3d}
is due to the discontinuous part of Eqs.~\eqref{G-BP-3d-t} and \eqref{G-BP-3d-grad} 
and they draw contribution emitted at all times $t'$ from $-\infty$ up to $t-R/c$. 

It can be seen that Eqs.~\eqref{E-ret-3d} and \eqref{B-ret-3d}
have some similarities but also differences to the so-called Jefimenko equations in Maxwell's electrodynamics~\citep{Jefimenko} 
(see also \citep{Lazar2013}).
The differences are based on the appearance of the Bopp-Podolsky Green function~\eqref{G-BP-3d} in the Bopp-Podolsky electrodynamics
instead of the Green function of the d'Alembert operator~\eqref{G-w-3d} in the Maxwell electrodynamics.

\subsubsection{2D}

In two-dimensional electrodynamics, the magnetic field strength is a scalar field $B_{(2)}=\nabla\times\bm A_{(2)}=\epsilon_{ij}\pd_i A_j$, where
$\epsilon_{ij}$ is the two-dimensional Levi-Civita tensor, and the electric
field strength $E_{(2)}=(E_x,E_y)$ is a two-dimensional vector field  
(see, e.g.,~\citep{Lap}).

Substituting the derivatives of the Bopp-Podolsky Green function~\eqref{G-BP-2d-t} 
and \eqref{G-BP-2d-grad} into Eqs.~\eqref{E-rp} and \eqref{B-rp}, 
the two-dimensional retarded electromagnetic field strengths become
\begin{align}
\label{E-ret-2d}
\bm E_{(2)}(\bm r,t)
&=-\frac{c}{2\pi \varepsilon_0}\int_{-\infty}^{t-R/c}\d t'\int_{\Bbb R^2}\d \bm r'\, 
\frac{\big[\bm R \rho(\bm r',t')-\tau \bm J(\bm r',t')\big]}{(c^2\tau^2-R^2)^\frac{3}{2}}\, 
\bigg[
1-\cos\bigg( \frac{\sqrt{c^2 \tau^2-R^2}}{\ell}\bigg)\nonumber\\
&\qquad
-\frac{\sqrt{c^2\tau^2-R^2}}{\ell}
\, \sin\bigg( \frac{\sqrt{c^2 \tau^2-R^2}}{\ell}\bigg)\bigg]
\end{align}
and
\begin{align}
\label{B-ret-2d}
B_{(2)}(\bm r,t)
&=\frac{\mu_0 c}{2\pi}\int_{-\infty}^{t-R/c}\d t'\int_{\Bbb R^2}\d \bm r'\, 
\frac{\bm R\times \bm J(\bm r',t')}{(c^2\tau^2-R^2)^\frac{3}{2}}\, 
\bigg[
1-\cos\bigg( \frac{\sqrt{c^2 \tau^2-R^2}}{\ell}\bigg)\nonumber\\
&\qquad
-\frac{\sqrt{c^2\tau^2-R^2}}{\ell}
\, \sin\bigg( \frac{\sqrt{c^2 \tau^2-R^2}}{\ell}\bigg)
\bigg] \,,
\end{align}
where
$\bm R\times \bm J=\epsilon_{ij} R_i J_j$.
The two-dimensional 
retarded electromagnetic field strengths~\eqref{E-ret-2d}
and \eqref{B-ret-2d} show an afterglow,
since they draw contribution emitted at all times $t'$ from $-\infty$ up to $t-R/c$.

\subsubsection{1D}

This version of the Bopp-Podolsky electrodynamics in one spatial dimension has a scalar electric
field and no magnetic field (see, e.g.,~\cite{Galic} for classical
electrodynamics in one spatial dimension). 

Substituting the derivatives of the Bopp-Podolsky Green function~\eqref{G-BP-1d-t} 
and \eqref{G-BP-1d-grad} into Eqs.~\eqref{E-rp} and \eqref{B-rp}, 
the one-dimensional retarded electromagnetic field strengths read as 
\begin{align}
\label{E-ret-1d}
E_{(1)}(x,t)
&=\frac{c}{2\varepsilon_0\ell}\int_{-\infty}^{t-|X|/c}
\d t'\int_{-\infty}^\infty\d x'\, 
\frac{\big[X\rho(x',t')-\tau J(x',t')\big]}{\sqrt{c^2\tau^2-X^2}}\,
J_1 \bigg( \frac{\sqrt{c^2\tau^2-X^2}}{\ell}\bigg)\,,
\\
\label{B-ret-1d}
B_{(1)}(x,t)
&=0\,.
\end{align}
The one-dimensional 
retarded electric field strength~\eqref{E-ret-1d}
possesses an afterglow,
because it draws contribution emitted at all times $t'$ from $-\infty$ up to $t-|X|/c$.

\section{Generalized Li{\'e}nard-Wiechert fields: 
electromagnetic fields of a non-uniformly moving point charge}
\label{sec5}

We consider 
a non-uniformly moving point charge carrying the charge $q$ 
at the position $\Bs(t)$.
The electric charge density and the electric current density vector are given by
\begin{align}
\label{J}
\rho(\rr,t)&=q \,\delta(\rr-\Bs(t))\,,\qquad
\BJ(\rr,t)=q \BV(t)\, \delta(\rr-\Bs(t))\,,
\end{align}
where $\BV(t)=\pd_t{\Bs}(t)=\dot{\Bs}(t)$ 
is the arbitrary velocity of the non-uniformly moving point charge.
We consider the case that the velocity of the point charge
is less than the speed of light: $|\BV|<c$. 
Therefore, the retarded potentials for non-uniformly moving point charges  lead to the 
generalized Li{\'e}nard-Wiechert potentials in the framework of Bopp-Podolsky electrodynamics.

\subsection{Generalized Li{\'e}nard-Wiechert potentials}

\subsubsection{3D} 

Substituting Eq.~(\ref{J}) into Eqs.~(\ref{phi-ret-3d}) and (\ref{A-ret-3d}) 
and performing the spatial integration,
the three-dimensional generalized Li{\'e}nard-Wiechert potentials read as 
\begin{align}
\label{phi-LW-3d}
\phi_{(3)}(\bm r,t)
&=\frac{q c}{4\pi\varepsilon_0\ell}\int_{-\infty}^\tR\d t'\,
\frac{1}{\sqrt{c^2 \tau^2-R^2(t')}}\, 
J_1 \bigg(\frac{\sqrt{c^2 \tau^2-R^2(t')}}{\ell}\bigg) 
\end{align}
and 
\begin{align}
\label{A-LW-3d}
\bm A_{(3)}(\bm r,t)
&=\frac{\mu_0 qc}{4\pi\ell}\int_{-\infty}^\tR\d t'\,
\frac{\bm V(t')}{\sqrt{c^2 \tau^2-R^2(t')}}\, 
J_1 \bigg(\frac{\sqrt{c^2 \tau^2-R^2(t')}}{\ell}\bigg)\,,
\end{align}
where $\bm R(t')=\bm r-\bm s(t')$
and the retarded time~$\tR$ being the root of the equation
\begin{align}
\label{tR-3d}
\big[x-s_x(\tR)\big]^2+\big[y-s_y(\tR)\big]^2+\big[z-s_z(\tR)\big]^2
-c^2(t-\tR)^2=0\,.
\end{align}
Due to the condition $|\BV|<c$, there is only one solution of Eq.~\eqref{tR-3d} which is the retarded time $\tR$.
The three-dimensional generalized Li{\'e}nard-Wiechert potentials~\eqref{phi-LW-3d}
and \eqref{A-LW-3d} draw contributions emitted at all times $t'$ from $-\infty$ up to $\tR$.
The generalized Li{\'e}nard-Wiechert potentials~\eqref{phi-LW-3d}
and \eqref{A-LW-3d} are in agreement with the expressions given by~\citet{Lande}.

\subsubsection{2D} 

Substituting Eq.~(\ref{J}) into Eqs.~(\ref{phi-ret-2d}) and (\ref{A-ret-2d}) 
and performing the spatial integration,
the two-dimensional generalized Li{\'e}nard-Wiechert potentials become
\begin{align}
\label{phi-LW-2d}
\phi_{(2)}(\bm r,t)
&=\frac{qc}{2\pi\varepsilon_0}\int_{-\infty}^\tR\d t'\,
\frac{1}{\sqrt{c^2\tau^2-R^2(t')}}\, 
\bigg[1-\cos\bigg( \frac{\sqrt{c^2 \tau^2-R^2(t')}}{\ell}\bigg)\bigg] 
\end{align}
and
\begin{align}
\label{A-LW-2d}
\bm A_{(2)}(\bm r,t)
&=\frac{\mu_0 qc}{2\pi}\int_{-\infty}^\tR\d t'\,
\frac{\bm V(t')}{\sqrt{c^2\tau^2-R^2(t')}}\, 
\bigg[1-\cos\bigg( \frac{\sqrt{c^2 \tau^2-R^2(t')}}{\ell}\bigg)\bigg]\,,
\end{align}
where $\bm R(t')=\bm r-\bm s(t')$
and the retarded time~$\tR$ being the root of the equation
\begin{align}
\label{tR-2d}
\big[x-s_x(\tR)\big]^2+\big[y-s_y(\tR)\big]^2-c^2(t-\tR)^2=0\,.
\end{align}
It can be seen that the two-dimensional generalized Li{\'e}nard-Wiechert potentials~\eqref{phi-LW-2d}
and \eqref{A-LW-2d} draw contributions emitted at all times $t'$ from $-\infty$ up to $\tR$. 

\subsubsection{1D} 

Substituting Eq.~(\ref{J}) into Eqs.~(\ref{phi-ret-1d}) and (\ref{A-ret-1d}), 
the spatial integration can be performed to
give the one-dimensional generalized Li{\'e}nard-Wiechert potentials
\begin{align}
\label{phi-LW-1d}
\phi_{(1)}(x,t)
&=\frac{qc}{2\varepsilon_0}\int_{-\infty}^\tR\d t'\, 
%%%H\big(c\tau-|X(t')|)\,
\bigg[1-J_0 \bigg( \frac{\sqrt{c^2\tau^2-X^2(t')}}{\ell}\bigg)\bigg]
\end{align}
and 
\begin{align}
\label{A-LW-1d}
A_{(1)}(x,t)
&=\frac{\mu_0 qc}{2}\int_{-\infty}^\tR\d t'\,   V(t')
%%%H\big(c\tau-|X(t')|)\,
\bigg[1-J_0 \bigg( \frac{\sqrt{c^2\tau^2-X^2(t')}}{\ell}\bigg)\bigg]\,,
\end{align}
where $X(t')=x-s(t')$
and $\tR$ is the retarded time, which is the root of the equation
\begin{align}
\label{tR-1d}
\big[x-s(\tR)\big]^2-c^2(t-\tR)^2=0\,.
\end{align}
Also the one-dimensional generalized Li{\'e}nard-Wiechert potentials~\eqref{phi-LW-1d}
and \eqref{A-LW-1d} draw contributions emitted at all times $t'$ from $-\infty$ up to $\tR$.

Like the retarded potentials in the Bopp-Podolsky electrodynamics, 
the generalized  Li{\'e}nard-Wiechert potentials in 3D, 2D and 1D,
Eqs.~(\ref{phi-LW-3d}), (\ref{A-LW-3d}), (\ref{phi-LW-2d}), (\ref{A-LW-2d})
and (\ref{phi-LW-1d}), (\ref{A-LW-1d})  depend on the entire history of the point charge up to
the retarded time $\tR$ and contain ``tail terms".

\subsection{Generalized Li{\'e}nard-Wiechert form of the electromagnetic field strengths}

\subsubsection{3D}

Substituting Eq.~(\ref{J}) into Eqs.~(\ref{E-ret-3d}) and (\ref{B-ret-3d}) 
and performing the spatial integration (see, e.g., \citep{Eyges,Jones86,Lazar2013}),
the three-dimensional electromagnetic fields in the 
generalized Li{\'e}nard-Wiechert form read as 
\begin{align}
\label{E-LW-3d}
\bm E_{(3)}(\bm r,t)
&=\frac{q}{8\pi\varepsilon_0 \ell^2}\,
\bigg[
\frac{\bm R(t')}{P(t')}-\frac{\bm V(t') R(t')}{c P(t')}\bigg]_{t'=\tR}
\nonumber\\
&\quad
-\frac{qc}{4\pi\varepsilon_0 \ell^2}\,
\int_{-\infty}^\tR\d t'\,
\frac{\bm R(t')-\tau \bm V(t')}{(c^2 \tau^2-R^2(t'))}\, 
J_2 \bigg(\frac{\sqrt{c^2 \tau^2-R^2(t')}}{\ell}\bigg)
%%\big[\bm R(t')-\tau \bm V(t')\big] 
\end{align}
and
\begin{align}
\label{B-LW-3d}
\bm B_{(3)}(\bm r,t)
&=-\frac{\mu_0 q}{8\pi\ell^2}\,
\bigg[\frac{\bm R(t')\times \bm V(t')}{P(t')}\bigg]_{t'=\tR}
\nonumber\\
&\quad 
+\frac{\mu_0 qc}{4\pi\ell^2}\,
\int_{-\infty}^\tR\d t'\, 
\frac{\bm R(t') \times \bm V(t') }{(c^2 \tau^2-R^2(t'))}\, 
J_2 \bigg(\frac{\sqrt{c^2 \tau^2-R^2(t')}}{\ell}\bigg)\,,
%% \bm R(t') \times \bm V(t') 
\end{align}
where
\begin{align}
\label{P}
P(t')=R(t')-\BV(t')\cdot\BR(t')/c\,.
\end{align}
In the first part of Eqs.~\eqref{E-LW-3d} and \eqref{B-LW-3d}, 
the expression inside the brackets has to be taken at the retarded time $t'=\tR$, which is the unique solution of Eq.~\eqref{tR-3d}.
The second part of Eqs.~\eqref{E-LW-3d} and \eqref{B-LW-3d}
draws contribution emitted at all times $t'$ from $-\infty$ up to the retarded time $\tR$.
Note that the term $\bm R(t')/P(t')$ in the first part of Eqs.~\eqref{E-LW-3d} and \eqref{B-LW-3d} possesses 
a (directional)  discontinuity (see also~\citep{Perlick2015}).

\subsubsection{2D}

Substituting Eq.~(\ref{J}) into Eqs.~(\ref{E-ret-2d}) and (\ref{B-ret-2d}) 
and performing the spatial integration,
the two-dimensional electromagnetic fields in the 
generalized Li{\'e}nard-Wiechert form become
\begin{align}
\label{E-LW-2d}
\bm E_{(2)}(\bm r,t)
&=-\frac{qc}{2\pi \varepsilon_0}\int_{-\infty}^\tR\d t'\,
\frac{\bm R(t')-\tau \bm V(t')}{(c^2\tau^2-R^2(t'))^\frac{3}{2}}\, 
\bigg[
1-\cos\bigg( \frac{\sqrt{c^2 \tau^2-R^2(t')}}{\ell}\bigg)\nonumber\\
&\qquad\qquad
-\frac{\sqrt{c^2\tau^2-R^2(t')}}{\ell}
\, \sin\bigg( \frac{\sqrt{c^2 \tau^2-R^2(t')}}{\ell}\bigg)\bigg]
\end{align}
and 
\begin{align}
\label{B-LW-2d}
B_{(2)}(\bm r,t)
&=\frac{\mu_0 qc}{2\pi}\int_{-\infty}^\tR\d t'\, 
\frac{\bm R(t')\times \bm V(t')}{(c^2\tau^2-R^2(t'))^\frac{3}{2}}\, 
\bigg[
1-\cos\bigg( \frac{\sqrt{c^2 \tau^2-R^2(t')}}{\ell}\bigg)\nonumber\\
&\qquad\qquad
-\frac{\sqrt{c^2\tau^2-R^2(t')}}{\ell}
\, \sin\bigg( \frac{\sqrt{c^2 \tau^2-R^2(t')}}{\ell}\bigg)
\bigg] \,.
\end{align}
It can be seen that the two-dimensional electromagnetic fields~\eqref{E-LW-2d}
and \eqref{B-LW-2d} draw contributions emitted at all times $t'$ from $-\infty$ up to $\tR$,
being the unique solution of Eq.~\eqref{tR-2d}.

\subsubsection{1D}

Substituting Eq.~(\ref{J}) into Eqs.~(\ref{E-ret-1d}) and (\ref{B-ret-1d}), 
the spatial integration can be performed to
give the one-dimensional electromagnetic fields in the 
generalized Li{\'e}nard-Wiechert form
\begin{align}
\label{E-LW-1d}
E_{(1)}(x,t)
&=\frac{qc}{2\varepsilon_0\ell}\int_{-\infty}^\tR\d t' \, 
\frac{ X(t')-\tau V(t') }{\sqrt{c^2\tau^2-X^2(t')}}\,
J_1 \bigg( \frac{\sqrt{c^2\tau^2-X^2(t')}}{\ell}\bigg)\,,\\
\label{B-LW-1d}
B_{(1)}(x,t)
&=0\,.
\end{align}
Thus, the one-dimensional electric field~\eqref{E-LW-1d}
draws contributions emitted at all times $t'$ from $-\infty$ up to $\tR$,
which is the unique solution of Eq.~\eqref{tR-1d}.

\section{Conclusion}
\label{sec6}

\begin{table}[t]
\caption{Behaviour of the Green function of the Bopp-Podolsky electrodynamics and its first derivatives on the light cone.}
\begin{center}
\leavevmode
\begin{tabular}{||c|c|c||}\hline
Spatial dimension & Green function $G^\text{BP}$ & First derivatives of $G^\text{BP}$\\
\hline
3D&  finite and discontinuous & singular and discontinuous \\
2D& approaching zero & singular and discontinuous \\
1D& approaching zero & finite and discontinuous\\
\hline
\end{tabular}
\end{center}
\label{table}
\end{table}

We have investigated the Bopp-Podolsky electrodynamics as prototype of 
a dynamical gradient theory with weak nonlocality in space and time.
The retarded potentials, retarded electromagnetic field strengths, generalized
Li{\'e}nard-Wiechert potentials and electromagnetic field strengths 
in generalized Li{\'e}nard-Wiechert form have been calculated for 3D, 2D and 1D and they depend 
on the entire history from $-\infty$ up to the retarded time $\tR$. 
The Bopp-Podolsky field is a superposition of the Maxwell field describing a massless photon
and the Klein-Gordon field describing a massive one.  
In particular, the Klein-Gordon part of the Bopp-Podolsky field gives rise to
a decreasing oscillation around the classical Maxwell field. 
The Green function of the Bopp-Podolsky electrodynamics and its first
derivatives have been calculated and studied in the neighbourhood 
of the light cone (see table~\ref{table}).  
It turned out that the Bopp-Podolsky Green function is the regularization of
the Green function of the d'Alembert operator:
\begin{align}
\label{reg}
G^\text{BP}=\text{reg}\, \big[G^\square\big]\,,
\end{align}
corresponding to the simplest case of the Pauli-Villars regularization 
with a single ``auxiliary mass'' proportional to $1/\ell$. 
The  Green function of the Klein-Gordon operator plays the mathematical role of the regularization function 
in the Bopp-Podolsky electrodynamics. 
Moreover, the retarded Bopp-Podolsky Green function and its first
derivatives show decreasing oscillations inside the forward light cone.
The behaviour of the electromagnetic potentials and electromagnetic field
strengths on the light cone is obtained from the behaviour of the Green
function and its first derivatives in the neighbourhood of the light
cone. Only in 1D the electric field strength of the Bopp-Podolsky
electrodynamics is singularity-free on the light cone. 
In 2D and 3D, the electromagnetic field strengths in the Bopp-Podolsky
electrodynamics possess weaker singularities than the classical singularities 
of the electromagnetic field strengths in the Maxwell electrodynamics. 
In order to regularize the 2D and 3D electromagnetic field strengths in 
the Bopp-Podolsky electrodynamics towards singular-free fields 
on the light cone, 
generalized electrodynamics of higher order might be used.
%% %and will be given in a forthcoming paper.

\section*{Acknowledgement}
The author gratefully acknowledges the grant from the 
Deutsche Forschungsgemeinschaft (Grant No. La1974/4-1).

\end{document}